\documentclass[11pt, fleqn]{paper}

\usepackage[T1]{fontenc}
\usepackage[utf8]{inputenc}

\usepackage[a4paper,margin=2.5cm]{geometry}
\usepackage{float, subcaption}

\usepackage{amsmath, amsfonts}
\usepackage{mathrsfs, upgreek, stmaryrd}
\usepackage{tensor, braket}
\allowdisplaybreaks

\usepackage[hidelinks]{hyperref}
\usepackage[capitalise]{cleveref}

\usepackage[backref=true,sorting=none,url=false,backend=biber,sortcites=true,style=numeric-comp]{biblatex}

\bibliography{bibliography}

\numberwithin{equation}{section}
\usepackage{mathtools}
\mathtoolsset{centercolon}
\usepackage[onehalfspacing]{setspace}

\usepackage{lmodern}
\usepackage[final]{microtype}

\def\EM{\textsc{em}}
\def\GW{\textsc{gw}}

\def\half{\tfrac{1}{2}}

\def\quarter{\tfrac{1}{4}}

\def\eighth{\tfrac{1}{8}}

\let\t\tensor
\let\p\partial
\let\wop\square
\def\dd{\mathrm{d}}
\def\Dd{\mathscr{D}}
\def\LieD{\mathfrak{L}}

\def\Ee{\mathbf{E}}
\def\Bb{\mathbf{B}}

\def\rNumbers{\mathbf R}

\def\o#1{{(#1)}}

\def\omegaG{\omega_g}

\def\cI{{\textsc{i}}}
\def\cII{{\textsc{ii}}}

\def\ee{\mathfrak e}
\def\ff{{\mathfrak{f}}}
\def\Aa{{\mathscr{A}}}
\def\Oo{{\mathscr{O}}}
\def\kc{{\check k}}
\def\RI{R_\cI}
\def\RII{R_\cII}
\def\Rbs{R_\text{BS}}
\def\Ff{{\mathscr F}}

\def\mI{{m^{}_\cI}}
\def\mII{{m^{}_\cII}}
\def\nuI{{\nu^{}_\cI}}
\def\nuII{{\nu^{}_\cII}}

\def\DCR{{\textsc{dcr}}}
\def\BHR{{\textsc{bhr}}}

\begin{document}
\title{The Response of Laser Interferometric Gravitational Wave Detectors Beyond the Eikonal Equation}
\author{\upshape Thomas B. Mieling\thanks{
	\textsc{orcid:} \href{https://orcid.org/0000-0002-6905-0183}{0000-0002-6905-0183},
	{email:} \href{mailto:thomas.mieling@univie.ac.at}{thomas.mieling@univie.ac.at}
	}
}
\date{\today}
\institution{\small University of Vienna, Faculty of Physics, TURIS Research Platform, Vienna, Austria}
\maketitle\vspace{-1.2em}
\noindent\today

\begin{abstract}
The response of Michelson interferometers to weak plane gravitational waves is computed at one order of accuracy beyond the eikonal equation.
The modulation of the electromagnetic field amplitude and polarisation are taken into account by solving the transport equations of geometrical optics with boundary conditions adapted to laser interferometry.
Considering both DC and balanced homodyne readout schemes, explicit formulae for the interferometer output signals are derived. These signals comprise perturbations of the optical path length, frequency and amplitude, and are shown to be insensitive to polarisation perturbations.
\end{abstract}

\tableofcontents

\section{Introduction}

Laser interferometric detectors of gravitational waves (\GW’s) have been studied extensively using a variety of methods. The perhaps simplest models are based on the Jacobi equation \cite{Forward1978, Schutz1987}, assuming the mirrors to follow nearby geodesics whose separation is read out interferometrically. The assumption of small arm lengths can be removed by computing round-trip times of light rays, which are commonly modelled either as null curves with prescribed spatial trajectories \cites{Weiss1972,Estabrook1975,Rakhmanov2008,Melissinos2010,Saulson2017} or as null geodesics \cites{Poplawski2006,Rakhmanov2009,Koop2014,Blaut2019}.
A comparison of these three methods can be found in Ref.~\cite{Finn2009}, where the underlying assumptions are assessed
and criteria are formulated under which circumstances these methods yield the same result.

While the ray round-trip time determines the optical phase via the eikonal equation, it does not provide information about \GW\ effects on the light amplitude \cite{Zipoy1966} and polarisation \cites{Skrotsky1957,Balazs1958}, the description of which requires methods beyond the eikonal equation, i.e.\ the Einstein-Maxwell equations or, in the high-frequency limit, the transport equations of geometrical optics.
The influence of gravity on electromagnetic (\EM) polarisation has become known as the gravitational Faraday effect and was studied extensively in various scenarios, i.a.\ for light coming from distant stars and being perturbed by the presence of massive objects \cites{Plebanski1960,Kopeikin2002,Sereno2004} or by gravitational waves, see e.g. Refs.~\cites{Codina1980,Lesovik2005,Faraoni_2008} and references therein. A recent analysis quantifying the \GW\ effect on light polarisation using Stokes parameters can be found in Ref.~\cite{Park2021}. Naturally, such calculations were mainly concerned with one-way light propagation and used emission conditions different from those suitable to describe laser interferometers.

To the best of our knowledge, the first description of laser interferometry beyond the eikonal equation was given in Ref.~\cite{Cooperstock1968} to compute the \GW\ response of standing \EM\ waves in Fabry-Pérot cavities. The analysis, however, was limited to special alignments, where the electromagnetic and gravitational waves propagate either in parallel or orthogonal directions. The general case of arbitrary alignment was discussed much later in Ref.~\cite{Lobo1992}. This calculation was criticised in Ref.~\cite{Cooperstock1993} for not implementing appropriate boundary conditions, but the alternative analysis presented there was again limited to a special alignment, where the \GW\ propagates orthogonally to both interferometer arms.
General alignments were reconsidered in Ref.~\cite{Montanari1998}, but the analysis of interferometers provided there was restricted again to the phase shift. Although giving explicit expressions for the full \EM\ field, the authors did not describe polarisation reflection at mirrors but instead considered other setups to determine the polarisation perturbation. The analysis was then extended to describe the \EM\ field in a cavity \cite{Calura1999}, but the applications were restricted to specific alignments and polarisations of the gravitational wave.

Subsequent analyses of \GW\ effects beyond the eikonal equation were provided for single light rays \cite{Cruise2000}, optical cavities \cite{Tarabrin2007} (where the problem is typically reduced to a single dimension for simplicity), and Ref.~\cite{Christie2007} provided examples of polarisation effects in Kerr-Schild \GW\ metrics. However, we are not aware of comprehensive analyses of laser interferometers taking all these effects into account and allowing for arbitrary alignments relative to the \GW.

The aim of this paper is to provide a systematic description of Michelson interferometers in the presence of weak \GW’s, based on the transport equations of geometrical optics. (A similar analysis based on the full Einstein-Maxwell equations is in preparation.) To this end, we model the emission of laser radiation, light propagation in interferometer arms, reflection at mirrors and the behaviour of light at beam splitters — all in the presence of a weak plane gravitational wave.
In particular, light emission is modelled by prescribing the field on a suitably chosen timelike hypersurface, where we impose the \EM\ field to have a definite frequency and constant intensity, but allow for general perturbations of the \EM\ polarisation. The reflection is assumed to take place at perfect mirrors, and the partial reflection at the beam splitter is described using a simple model which is motivated by the description of light reflection at perfectly reflecting surfaces.
The resulting expression for the light intensity at the detector, which is derived from the energy-momentum tensor, thus contains \GW\ effects on the optical phase, amplitude and polarisation. The calculations presented here allow for arbitrary incidence angles, polarisations and waveforms of the gravitational wave and do not involve unidimensional approximations.

The structure of this paper is as follows. In \cref{s:michelson layout} we describe the setup of simple Michelson interferometers used for gravitational wave detection, and in \cref{s:transport equations} we review the fundamental equations of geometrical optics. The emission of monochromatic plane waves by a laser source and the propagation of such radiation in the \GW\ background is described in \cref{eq:michelson emission}. Reflection of such radiation by perfect mirrors is modelled in \cref{s:michelson mirror}, leading to a description of light rays as they travel back and forth in an interferometer arm. In \cref{s:michelson beam splitter} we give a model for the behaviour of the fields at (non-polarising) beam splitters, allowing to compute the radiation sent into the two arms from the laser source, and to compute the radiation reaching the detector via the output port. The results are combined in \cref{s:michelson output}, where the output intensity of the interferometer is computed and discussed. Finally, approximate formulae for the interferometer output are derived and plotted in \cref{s:pattern functions}.

After finalising this paper, we became aware of the recent preprint \cite{Lobato2021}, where the same setup is analysed.

\section{Setup: Michelson Interferometers and Plane Gravitational Waves}
\label{s:michelson layout}

Consider a linearised plane gravitational wave of amplitude $\varepsilon \ll 1$ in transverse, traceless, and synchronous gauge
\begin{equation}
	\label{eq:metric TT}
	\t g{_\mu_\nu}
		= \t \eta{_\mu_\nu}
		+ \varepsilon \t h{_\mu_\nu}(\t \kappa{_\rho} \t x{^\rho})\,,
\end{equation}
where $\eta$ is the background Minkowski metric (with the sign convention that spacelike vectors have positive norm), and $\kappa$ is the \GW\ wave vector
\begin{equation}
	\t \kappa{_\mu} = \omegaG (1, - \t n{_i})\,,
\end{equation}
where $\omegaG$ is the \GW\ frequency and $\t n{_i}$ is a unit vector in the unperturbed geometry. Note that we take $\kappa$ to be \emph{past pointing} so that for a wave propagating along the $x$-axis, the argument $\t \kappa{_\rho} \t x{^\rho}$ coincides (up to the factor $\omegaG$) with the standard null coordinate $u = t - x$. The gauge conditions then require
\begin{align}
	\t h{_\mu_0} &= 0\,,
	&
	\t h{_\mu^\nu} \t \kappa{_\nu} &= 0\,,
	&
	\t h{^\mu_\mu} &= 0\,.
\end{align}
i.e.\ the metric perturbation is purely spatial, transverse and traceless (where indices of $h$ are raised and lowered with the background Minkowski metric).

This paper uses the following notation. For any two vectors $v$ and $w$ we write $v.w \equiv \t \eta{_\mu_\nu} \t v{^\mu} \t w{^\nu}$, and similarly we write the argument of the metric perturbation as $\kappa.x \equiv \t \kappa{_\mu} \t x{^\mu}$. Further, for quadratic forms such as $h$ and $g$ we use the notation $h(v,w) = \t h{_\mu_\nu} \t v{^\mu} \t w{^\nu}$.

Let us now consider Michelson interferometers in the metric \eqref{eq:metric TT}.
Here, we follow the standard practice of describing the geometry of the interferometer in terms of the (unperturbed) background metric. Physically, this can be justified by assuming the apparatus to be aligned before the \GW\ arrives, and assuming all material points of the interferometer to follow geodesic motion as the \GW\ passes by. For extended bodies, such as the beam splitter or the mirrors, deviations from this model could be described using elasticity theory, but this is beyond the scope of this paper. Note, however, that such arising corrections to the boundary conditions will be of order $\varepsilon$ and can thus be described using the \emph{unperturbed} equations of geometrical optics.

\begin{figure}[ht]
	\centering
	\includegraphics[width=8cm]{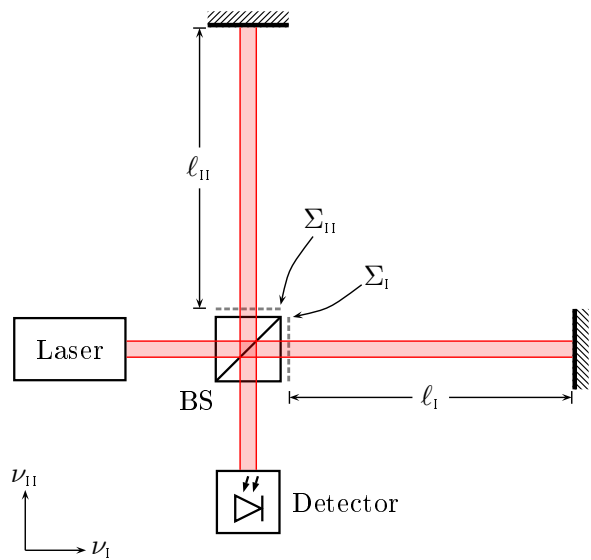}
	\caption{Schematics of a Michelson interferometer with the two emission surfaces $\Sigma_\cI$ and $\Sigma_\cII$ behind the beam splitter (BS) indicated by dashed lines.}
	\label{fig:michelson}
\end{figure}

\Cref{fig:michelson} shows a simplified schematic drawing of a Michelson interferometer. (The current \textsc{ligo} and Virgo interferometers use additional recycling mirrors and Fabry-Pérot cavities, see e.g.\ Ref.~\cite[Sect.~5]{Pitkin2011}, but here we restrict ourselves to this simplified setup.) In typical applications for gravitational wave detection, the distances between the laser, beam splitter, and detector are much shorter than the two arm lengths of the interferometer. Hence, we will neglect \GW\ effects in the segments {(laser $\to$ beam splitter)} and {(beam splitter $\to$ detector)}, and focus on the propagation of light in the two interferometer arms.

Accordingly, we describe the radiation sent into the two interferometer arms by boundary values for the electromagnetic field on the two surfaces $\Sigma_\cI = \{ \mI . x = 0 \}$ and $\Sigma_\cII = \{ \mII . x = 0 \}$ (cf.\ \cref{fig:michelson}), where $\mI$ and $\mII$ are orthogonal unit vectors in the background geometry. The radiation then propagates towards the mirrors placed at $\mI . x = \ell_\cI$ and $\mII . x = \ell_\cII$, reflects there and propagates back to the emission surfaces $\Sigma_\cI$ and $\Sigma_\cII$. The propagation of these light rays will be described by the equations of geometrical optics, which are reviewed in the next section. We solve these equations for arbitrary polarisation prescribed on $\Sigma_\cI$ and $\Sigma_\cII$ and then specialise to physical boundary values derived from a model of the beam splitter, which is necessary to describe how the laser light is transferred into the arms and how the returning beams are transferred towards the detector. The various intermediate results are then put together in \cref{s:michelson output} to determine the output signal of the \GW\ detector.

\section{Review of Geometrical Optics}
\label{s:transport equations}

This section briefly reviews the transport equations of geometrical optics in a notation useful for the following analysis. The equations presented here coincide with those of Ref.~\cite{Christie2007}, see also Ref.~\cite{Christie2007a} for the relation to other formalisms based on 3+1 decompositions such as those of Refs.~\cites{Dehnen_1973,Santana_2020}.

Geometrical optics at leading order\footnote{There seems to be no universally agreed-upon nomenclature for counting orders in geometrical optics: the methods used here correspond to “order zero” in the scheme of Refs.~\cites{Kravtsov1990,Perlick2000}, while Ref.~\cite{Montanari1998} describes the same level of accuracy as “one order of approximation beyond the limit of geometrical optics.”} describes the electromagnetic field strength in the form%
\begin{equation}
	F = \ff\, e^{i \psi}\,,
\end{equation}
or sums of such expressions, where the eikonal $\psi$ satisfies the eikonal equation
\begin{equation}
	\label{eq:eikonal equation}
	\t g{^\mu^\nu} (\t \nabla{_\mu} \psi) (\t \nabla{_\nu} \psi) = 0\,,
\end{equation}
and the amplitude $\ff$ (a two-form) satisfies
\begin{align}
	\t \ff{_\mu_\nu} \t \nabla{^\mu} \psi &= 0\,,
	&
	\t \ff{_[_\mu_\nu} \t \nabla{_\rho_]} \psi &= 0\,,
	&
	\t \nabla{^\sigma} \psi \, \t \nabla{_\sigma} \t \ff{_\mu_\nu}
	+ \half (\wop \psi) \t \ff{_\mu_\nu} &= 0\,.
\end{align}
A motivation for these equations and the extension to higher order expansions is sketched in Appendix \ref{appendix:geometrical optics}.
Writing the amplitude as
\begin{equation}
	\t \ff{_\mu_\nu} = \Aa \t f{_\mu_\nu}\,,
\end{equation}
where $\mathscr A$ satisfies the scalar amplitude transport equation
\begin{equation}
	\label{eq:amplitude transport}
	\t \nabla{^\sigma} \psi \, \t \nabla{_\sigma} \Aa + \half (\wop \psi) \Aa = 0\,,
\end{equation}
the equations for the \EM\ polarisation tensor $\t f{_\mu_\nu}$ reduce to
\begin{align}
	\label{eq:polarisation constraint + transport}
	\t f{_\mu_\nu} \t \nabla{^\mu} \psi &= 0\,,
	&
	\t f{_[_\mu_\nu} \t \nabla{_\rho_]} \psi &= 0\,,
	&
	\t \nabla{^\sigma} \psi \, \t \nabla{_\sigma} \t f{_\mu_\nu} &= 0\,,
\end{align}
so that $\t f{_\mu_\nu}$ is parallel transported along the rays (integral curves of $\nabla \psi$). This system of equations is consistent because $\nabla \psi$ satisfies the geodesic equation.
The first two equations here entail that $f$ can be written as
\begin{equation}
	\label{eq:polarisation f = k E}
	f = (\dd \psi/\omega) \wedge E\,,
\end{equation}
where the \EM\ polarisation one-form $E$ is orthogonal to $\dd \psi$ (and one is always free to add any multiple of $\dd \psi$). The factor $1/\omega$ was introduced here so that $f$ has the same units as $E$.
As made precise in \cref{s:polarisation interpretation}, the field $E$ encodes the polarisation of the electromagnetic wave. Solutions to the system \eqref{eq:polarisation constraint + transport} are readily constructed by choosing $E$ to be parallel transported along the rays, i.e. $\nabla_{\nabla \psi} E = 0$, which is consistent with the condition $\t E{_\mu} \t \nabla{^\mu} \psi = 0$.
Since the norm of $E$ is preserved under parallel transport, the amplitude of the field is fully characterised by $\Aa$, so that one may assume $E$ to be normalised according to $g(E, E) = 1$. ($E$ is necessarily spacelike, since it is orthogonal to the lightlike vector $\nabla \psi$ and linearly independent of it.)
We are thus led to the system of equations
\begin{align}
	\label{eq:E transport}
	\t \nabla{^\sigma} \psi \, \t \nabla{_\sigma} \t E{_\mu} &= 0\,,
	&
	\t E{_\mu} \t \nabla{^\mu} \psi &= 0\,,
	&
	\t g{^\mu^\nu} \t E{_\mu} \t E{_\nu} &= 1\,.
\end{align}
Because $\nabla \psi$ is geodetic, the last two conditions propagate, i.e.\ if they hold at any point of a light ray, they hold along the entire ray.

The geometrical optics description of the field (at leading order) is thus obtained by first solving the eikonal equation \eqref{eq:eikonal equation}, and then determining the amplitude $\Aa$ from \eqref{eq:amplitude transport}, and the polarisation $E$ from \eqref{eq:E transport}. In all cases, one must prescribe boundary values on a non-characteristic hypersurface $\Sigma$. In the next section, we prescribe such boundary data as to model the emission of light by a monochromatic source  and solve for the eikonal, amplitude and polarisation.

\section{Emission of Monochromatic Plane Waves}
\label{eq:michelson emission}

To model the emission of light from a radiating surface, we prescribe boundary data on a \emph{timelike} hypersurface $\Sigma$, for which we take as the simplest model a coordinate-hyperplane
\begin{align}
	\Sigma &= \{ m.x = 0\}\,,
	&
	\text{where }
	\t m{_\mu} &= (0, \t m{_i})\,.
\end{align}
Here, the $\t m{_i}$’s are constants which may be taken to be normalised with respect to the unperturbed spatial metric, i.e.\ $\t \delta{^i^j} \t m{_i} \t m{_j} = 1$, so that $m.m = 1$.
The unit conormal (normal one-form) to this surface is given by
\begin{equation}
	\label{eq:normal perturbed}
	\nu = m \left( 1 + \half \varepsilon\, h(m,m) \right)\,.
\end{equation}
For the eikonal, we take $\psi = - \omega t$ on $\Sigma$ (corresponding to coherent emission with frequency $\omega$). This entails that on $\Sigma$ we have $\dd \psi = - \omega \dd t + \alpha \nu$, where $\alpha$ is determined by the eikonal equation up to a sign: $\alpha = \pm \omega$. We shall assume $\nu(\psi)|_\Sigma > 0$, corresponding to emission along $\nu$ (instead of emission in the opposite direction).

For the amplitude $\Aa$ and the polarisation $E$, however, one cannot directly carry over the unperturbed boundary data to the perturbed problem: as we will see, the boundary data on $\Sigma$ must satisfy a set of equations (specifically, the algebraic conditions of \cref{eq:E transport} and certain constraint equations discussed below) which differ from those of flat space, so that perturbations of the boundary data are unavoidable.

The algebraic conditions state that the prescribed polarisation, $E$, on $\Sigma$ must be orthogonal to $\dd \psi = \omega (\nu - \dd t)$, and normalised (both in the perturbed metric), as dictated by \cref{eq:E transport}.
Without loss of generality, $E$ can be chosen to be purely spatial on $\Sigma$ (this can be accomplished by adding to $E$ a suitable multiple of $\dd \psi$, cf.\ \cref{s:polarisation interpretation}), so that the algebraic conditions reduce to
\begin{align}
	\t E{_0}|_\Sigma &= 0\,,
	&
	g(E, \nu)|_\Sigma &= 0\,,
	&
	g(E, E)|_\Sigma &= 1\,.
\end{align}
Seeking solutions of the form $E = E^\o0 + O(\varepsilon)$, one obtains the perturbed polarisation on $\Sigma$
\begin{equation}
	\label{eq:emitted polarisation}
	E|_\Sigma
		= E^\o0
		+ \varepsilon m\, h(m, E^\o0)
		+ \half \varepsilon E^\o0\, h(E^\o0, E^\o0)
		+ \varepsilon E'\,,
\end{equation}
where
\begin{align}
	\label{eq:E' properties}
	\t{{E'}}{_0} &= 0\,,
	&
	m . E' &= 0\,,
	&
	E^\o0 . E' &= 0\,.
\end{align}
The first correction term in \cref{eq:emitted polarisation} ensures that $E$ is orthogonal to $\nu$ in the perturbed geometry, the second correction term fixes the normalisation, and the last term allows for small variations of the emitted polarisation.

Additionally, as shown in \Cref{appendix:constraints}, the emitted polarisation must satisfy certain differential constraints, analogous to the divergence constraints in the usual spacelike Cauchy problem. Denoting by ${\tilde E}$ the pull-back of the overall field $\Aa E$ to $\Sigma$ at any instant of time, they read
\begin{align}
	\label{eq:polarisation constraint equations}
	\t{\tilde \nabla}{_A} \t{\tilde E}{^A} &= 0\,,
	&
	\t{\tilde\epsilon}{^A^B} \t{\tilde \nabla}{_A} \t{\tilde E}{_B} &= 0\,,
\end{align}
where $A,B$ are spatial indices associated to the surface $\Sigma$ (at any instant of time), ${\tilde \nabla}$ is the induced Levi-Civita connection, and $\tilde \epsilon$ denotes the induced volume form.

In flat space, these equations imply that the components $\t{\tilde E}{_A}$ are harmonic functions, so that the requirement that the field be bounded implies that ${\tilde E}^\o0$ is spatially constant on $\Sigma$. Since one is free to prescribe the time dependence of ${\tilde E}^\o0$, the solutions are parameterised by two real functions of time alone. Plane polarised waves of constant energy density then correspond to $\Aa^\o0 = 1$ and $E^\o0$ being a constant spatial vector of unit norm, tangent to $\Sigma$.

In the perturbed case, the solutions of \cref{eq:polarisation constraint equations} are still parameterised by two real functions of time $t$ alone, but contrary to the unperturbed case it is not possible in general to arrange for $\tilde E$ to have constant norm everywhere on $\Sigma$ (see \cref{appendix:constraints}).
However, it turns out that the precise details of the perturbations are not relevant for the applications considered here, as one can nonetheless impose $\t{\tilde E}{^A}\t{\tilde E}{_A} = 1$ to hold at the spatial coordinate origin with significant deviations arising only at length scales comparable to the  \GW\ wavelength. Hence, as the field is only read out on much smaller regions, such perturbations will not contribute to the final output signal.

Consistent boundary data can thus be obtained by construction a solution $\t{\tilde E}{_A}$ of the form $\t*{\tilde E}{^{\o0}_A} + \varepsilon \t*{\tilde E}{^{\o1}_A}$ to \cref{eq:polarisation constraint equations} (using the methods described in \cref{appendix:constraints}) and separating amplitude and polarisation information by writing the perturbation in the form
\begin{equation}
	\t*{\tilde E}{^{\o1}_A}
		= \left[ \half h({\tilde E}^{\o0}, {\tilde E}^\o0) + \delta \Aa \right] \t*{\tilde E}{^{\o0}_A}
		+ \t{\tilde E{'}}{_A}\,,
\end{equation}
where ${\tilde E}'$ is orthogonal to ${\tilde E}^\o0$ in the background metric.
$\delta \Aa$ then corresponds to the amplitude perturbation, and extending the form ${\tilde E}'$ trivially to a space-time object by defining the temporal and normal components to be zero, one obtains a polarisation of the form \eqref{eq:emitted polarisation}, which satisfies all constraints.

We stress that $\delta \Aa$, ${\tilde E}$, and thus also $E'$ are defined on $\Sigma$ only (the propagation of the fields away from this surface will be computed later). To emphasise this point and to simplify the notation in the following calculations, we write
\begin{equation}
	\delta \Aa = \delta \Aa(\tau.x, \xi.x, \zeta.x)\,,
\end{equation}
and similarly for ${\tilde E}$ and $E'$, where $\tau, \xi, \zeta$ and $m$ form an orthonormal basis in the unperturbed geometry (for example, one could take $\tau = \dd t$ and $(m, \xi, \zeta)$ an orthonormal basis of $\rNumbers^3$).

With this notation, the boundary conditions for the amplitude and polarisation can be written as
\begin{align}
	\label{eq:boundary values A and E}
	\Aa|_\Sigma &= 1 + \varepsilon \delta \Aa(\tau.x, \xi.x, \zeta.x)\,,
	&
	E|_\Sigma &= E^\o0 + \varepsilon \delta E(\tau.x, \xi.x, \zeta.x)\,,
\end{align}
where we find it useful to decompose the polarisation perturbation into normal and tangential parts according to
\begin{align}
	\label{eq:polarisation perturbation decomposition}
	\delta E
		&= \delta E^\perp + \delta E^\parallel\,,
	&
	\delta E^\perp
		&= m\, h(m, E^\o0)\,,
	&
	\delta E^\parallel
		&= \half E^\o0 h(E^\o0, E^\o0) + E'\,.
\end{align}
As mentioned above, the precise details of the functions $\delta\Aa$ and $E'$ are not important for the analysis here: the only assumption on the amplitude is
\begin{equation}
	\label{eq:amplitude correction origin}
	\delta\Aa(t, \t x{^i} = 0) = 0\,,
\end{equation}
so that the field at the emission point has unit energy density, and the assumptions on $E'$ are to be found in \cref{eq:E' properties}.

To summarise, we consider the equations of geometrical optics with the following data prescribed on the surface $\Sigma = \{m.x = 0\}$:
\begin{align}
	\label{eq:emission bdry eikonal}
	\psi|_\Sigma
	&= - \omega t\,,
	\\
	\label{eq:emission bdry amplitude}
	\Aa|_\Sigma
	&= 1 + \varepsilon \delta \Aa(\tau.x, \xi.x, \zeta.x)\,,
	\\
	\label{eq:emission bdry polarisation}
	E|_\Sigma
		&= E^\o0 + \varepsilon \delta E(\tau.x, \xi.x, \zeta.x) \,,
\end{align}
together with the condition $\nu(\psi)|_\Sigma > 0$.

\paragraph{Eikonal Equation}

In the unperturbed case, the eikonal is
\begin{align}
	\psi^\o0
		&= k.x \equiv \t k{_\mu} \t x{^\mu}\,,
	& \text{where }
	\t k{_\mu} &= \omega(-1, \t m{_i})\,.
\end{align}
For the perturbed eikonal, the ansatz $\psi = k.x + \varepsilon \psi^\o1$ leads to
\begin{equation}
	\t k{^\mu} \t \p{_\mu} \psi^\o1
		= \half \t h{^\mu^\nu}(\kappa.x) \t k{_\mu} \t k{_\nu}\,,
\end{equation}
which is solved by integrating the right-hand side along the unperturbed rays.
To this end, we write $\t x{^\mu} = \t*x{_0^\mu} + s \t k{^\mu}$, where $\t*x{_0^\mu}$ lies on the emission surface $\Sigma$. Contracting with $\t m{_i}$, one finds $s = m.x/\omega$, so that the emission point (for any given $x$ away from $\Sigma$) is seen to be
\begin{equation}
	\label{eq:geometric x0}
	\t* x{_0^\mu}(x) = \t x{^\mu} - (m.x) \t k{^\mu} / \omega\,.
\end{equation}
Note that by construction one has $\t k{^\nu} \t \p{_\nu} \t* x{_0^\mu} = 0.$
Defining the integrated waveform
\begin{equation}
	\label{eq:scalar propagator}
	H(k, u_2, u_1)
		:= \frac{\t k{_\mu} \t k{_\nu}}{2 \kappa.k} \int_{u_1}^{u_2} \t h{^\mu^\nu}(u)\, \dd u\,,
\end{equation}
one obtains the eikonal
\begin{equation}
	\label{eq:eikonal solution 1}
	\psi
		= k.x
		+ \varepsilon H(k, \kappa.x, \kappa.x_0) \,.
\end{equation}
This clearly satisfies the boundary conditions: since $\kappa.x_0$ coincides with $\kappa.x$ on $\Sigma$, the first-order correction vanishes there.

Taking the derivative, the local wave vector is found to be
\begin{equation}
	\label{eq:eikonal gradient}
	\dd \psi
		= \omega (\nu - \dd t)
		+ \varepsilon [\kappa + m (\kappa.k/\omega)] \frac{h(k, k)}{2 \kappa.k} \bigg|_{\kappa.x_0}^{\kappa.x}
		\,,
\end{equation}
where $\nu$ is as in \cref{eq:normal perturbed} (but evaluated away from the emission surface $\Sigma$), and where we have used the notation
\begin{equation}
	h(k, k) \big|_{\kappa.x_0}^{\kappa.x}
	= \t k{_\mu} \t k{_\nu} \left(
		\t h{^\mu^\nu}(\kappa.x)
		- \t h{^\mu^\nu}(\kappa.x_0)
	\right)\,.
\end{equation}

\paragraph{Amplitude Transport Equation}

Consider now the transport equation \eqref{eq:amplitude transport} for the scalar amplitude $\Aa$.
In the absence of a gravitational wave (unperturbed case), a plane wave has $\Aa = 1$, so for the perturbed case we write $\Aa = 1 + \varepsilon \Aa^\o1$ and obtain at first order
\begin{align}
	\label{eq:A1 transport eq}
	\t k{^\mu} \t \p{_\mu} \Aa^\o1 &= - \half (\wop \psi)^\o1\,,
	\\
	\label{eq:A1 transport bdry}
	\Aa^\o1 |_\Sigma &= \delta \Aa(\tau.x, \xi.x, \zeta.x)\,.
\end{align}
Since we are working in harmonic coordinates, the scalar wave operator is simply $\wop = \t g{^\mu^\nu} \t \p{_\mu} \t \p{_\nu}$. Because the first term in \cref{eq:eikonal solution 1} has vanishing second derivatives, only the second term contributes, where the unperturbed expression for the wave operator suffices.
Using ${\t \p{_\mu} (\kappa.x) = \t \kappa{_\mu}}$ and $\t \p{_\mu} (\kappa.x_0) = \t \kappa{_\mu} - (\kappa.k/\omega) \t m{_\mu}$, one finds
%
\begin{equation}
\begin{split}
	\wop H(k, \kappa.x, \kappa.x_0)
		&= - \half (\kappa.k/\omega^2 - 2 \kappa.m / \omega) \t {{h'}}{^\rho^\sigma}(\kappa.x_0) \t k{_\rho} \t k{_\sigma}
		\\
		&= - \tfrac{\omegaG}{2 \omega} (1 + m.n) \t {{h'}}{^\rho^\sigma}(\kappa.x_0) \t k{_\rho} \t k{_\sigma}\,,
\end{split}
\end{equation}
where $\t {{h'}}{^\rho^\sigma}(u) = \t \p{_u} \t {h}{^\rho^\sigma}(u)$.
Consequently, the transport equation for $\Aa^\o1$ takes the form
\begin{equation}
	\t k{^\mu} \t \p{_\mu} \Aa^\o1
	= \frac{\omegaG}{4 \omega} (1 + m.n) \t {{h'}}{^\rho^\sigma}(\kappa.x_0) \t k{_\rho} \t k{_\sigma}\,.
\end{equation}
Since $\t k{^\mu} \t \p{_\mu}(\kappa.x_0) = 0$, this leads to an amplitude which grows linearly with the distance from the emission surface. Indeed, imposing the boundary condition \eqref{eq:A1 transport bdry}, the solution is
\begin{align}
	\Aa^\o1
		= \quarter (1 + m.n) \omegaG m.x\, \t {{h'}}{^\rho^\sigma}(\kappa.x_0) \t k{_\rho} \t k{_\sigma}/\omega^2
		+ \delta \Aa(x_0)\,.
\end{align}
The second term here, $\delta \Aa(x_0)$, corresponds to the boundary value perturbations of \cref{eq:A1 transport bdry}, evaluated at the emission point $x_0(x)$, as defined in \cref{eq:geometric x0}.
Due to the linear growth of the first term, we expect the leading order geometrical optics approximation to be only applicable for short distances from the emission surface $\Sigma$.

\paragraph{Polarisation Transport Equation}

In the unperturbed case, $\t E{^{\o0}_\mu}$ is a constant vector on $\Sigma$ with
\begin{align}
	\label{eq:polarisation constraints}
	\t E{^{\o0}_0} &= 0\,,
	&
	m.E^\o0 &= 0 \,,
\end{align}
i.e.\ the emitted polarisation is purely spatial and orthogonal to $m$. The transport equation then implies that $\t E{^{\o0}_\mu}$ is constant everywhere (also away from $\Sigma$).

In the perturbed case, we write $E = E^\o0 + \varepsilon E^\o1$ and impose on $E$ the boundary values given in \cref{eq:emission bdry polarisation}.
At first order, the equations for the polarisation are thus
\begin{align}
	\t k{^\sigma} \t \p{_\sigma} \t E{^{\o1}_\mu}
		&= \t \Gamma{^{\o1}^\nu_\mu_\rho} \t k{^\rho} \t E{^{\o0}_\nu} \,,
	\\
	\label{eq:polarisation boundary values general}
	\t E{^{\o1}_\mu}|_{\Sigma}
		&= \t{\delta E}{_\mu}(\tau.x, \xi.x, \zeta.x)\,,
\end{align}
where $\t \Gamma{^{\o1}^\nu_\mu_\rho}$ denotes the Christoffel symbols at first order in $\varepsilon$.
These equations are readily integrated to yield
\begin{align}
	\t E{^{\o1}_\mu}(x)
	&= \t{\varGamma(k,\kappa.x, \kappa.x_0)}{^\nu_\mu} \t E{^{\o0}_\nu}
	+ \t {\delta E}{_\mu}(x_0)\,,
\shortintertext{where}
	\label{eq:tensor propagator}
	\t {\varGamma(k, u_2, u_1)}{^\mu_\nu}
		&= \frac{1}{\kappa.k} \int_{u_1}^{u_2} \t \Gamma{^\mu_\nu_\rho}(u) \t k{^\rho}\, \dd u\,.
\end{align}
As for the scalar amplitude, $\t {\delta E}{_\mu}(x_0)$ corresponds to boundary values \eqref{eq:polarisation boundary values general} evaluated at the emission point $x_0(x)$, as defined in \cref{eq:geometric x0}.

Explicitly, using the standard formula for the Christoffel symbols, $\varGamma(k, u_2, u_1)$ is found to be given by
\begin{equation}
	\label{eq:Christoffel symbols integrated}
	\t {\varGamma(k, u_2, u_1)}{^\mu_\nu}
		= \frac{1}{2 \kappa.k} \t k{^\rho} \left(
			\t \kappa{_\rho} \t{{h}}{^\mu_\nu}(u)
			+ \t \kappa{_\nu} \t{{h}}{^\mu_\rho}(u)
			- \t \kappa{^\mu} \t{{h}}{_\nu_\rho}(u)
		\right)_{u=u_1}^{u=u_2}\,.
\end{equation}

\paragraph{Summary}
All in all, the leading order geometrical optics approximation for the electromagnetic field satisfying the emission conditions \eqref{eq:emission bdry eikonal}—\eqref{eq:emission bdry polarisation} was found to be
\begin{align}
	\label{eq:result emission}
	F
		&= \Aa e^{i \psi} \, (\dd \psi/\omega) \wedge E\,,
	\\
\shortintertext{with}
	\psi
		&= k.x
		+ \varepsilon H(k, \kappa.x, \kappa.x_0)\,,
		\\
	\Aa
		&= 1
		+ \quarter \varepsilon (1 + m.n) \omegaG m.x\, h'(m, m)_{\kappa.x_0}
		+ \varepsilon \delta \Aa(x_0)\,,
	\\
	\label{eq:result polarisation}
	\t E{_\mu}
		&= \t E{^{\o0}_\mu}
		+ \varepsilon \t{\varGamma(k, \kappa.x, \kappa.x_0)}{^\nu_\mu} \t E{^{\o0}_\nu}
		+ \varepsilon \t {\delta E}{_\mu}(x_0)
		\,,
\end{align}
where $m$ and $n$ are unit vectors specifying the propagation direction of the electromagnetic and gravitational waves, respectively, and where the subscript of $h$ indicates the value of the parameter $u$ where the metric perturbation is evaluated.

\section{Reflection at Mirrors}
\label{s:michelson mirror}

Consider the reflection of the wave \eqref{eq:result emission} at a mirror described by the timelike hypersurface
\begin{equation}
	\tilde\Sigma = \{ m.x = \ell \}\,,
\end{equation}
i.e.\ a surface which — in our coordinate system — is parallel to the emission surface $\Sigma$ and separated from it by a coordinate distance $\ell$.

As the simplest model, we assume the mirror to be perfectly reflecting. (More realistic models would use generalisations of the Fresnel equations to the general relativistic setting, as described in Ref.~\cite{Kremer1967}.)
As explained in \cref{appendix:perfect reflection}, the corresponding boundary conditions relating the reflected field $\check F = \check \Aa \check f e^{i \check \psi}$ to the incident field $F = \Aa f e^{i \psi}$ are
\begin{align}
	\label{eq:reflection conditions}
	\check \psi |_{\tilde\Sigma}
		&= \psi |_{\tilde\Sigma}\,,
	&
	\check \Aa |_{\tilde\Sigma}
		&= \Aa |_{\tilde\Sigma}\,,
	&
	\t{\check f}{_\alpha_\beta}|_{\tilde\Sigma}
		&= -\, \t R{_\alpha^\rho} \t R{_\beta^\sigma} \t f{_\rho_\sigma}|_{\tilde\Sigma}\,,
\end{align}
where $\t R{_\alpha^\beta}$ is the reflection along the surface normal to $\nu$ as defined in \cref{eq:normal perturbed}:
\begin{equation}
	\label{eq:reflection tensor}
	\t R{_\alpha^\beta}
		= \t*\delta{_\alpha^\beta}
		- 2 \t \nu{_\alpha} \t \nu{^\beta}\,,
\end{equation}
where the index of $\nu$ is raised with the full metric $g$.
These equations say that the phase and amplitude of the reflected wave coincide with those of the incident wave, and that the reflected polarisation is obtained from the incident polarisation by applying the reflection operator $R$ and changing the sign. (One could, alternatively, include the negative sign in $\check \Aa$, or add $\pm \pi$ to the reflected phase, but this does not alter the overall field $\check F$.)

To describe the reflected eikonal, we set $\kc$ to be the unperturbed reflected wave vector
\begin{equation}
	\t \kc{_\mu}
		= \t R{^{\o0}_\mu^\nu} \t k{_\nu}
		= \omega(-1, - \t m{_i})\,,
\end{equation}
and analogously to the definition of $x_0$ in \cref{eq:geometric x0}, let
\begin{equation}
	\label{eq:geometric x1}
	\t* x{^\mu_1}(x)
		= \t x{^\mu} - (\ell - m.x) \t \kc{^\mu}/\omega\,.
\end{equation}
With this notation, the eikonal of the reflected wave is readily found to be
\begin{equation}
	\check \psi
		= 2 \omega \ell
		+ \kc.x
		+ \varepsilon H(\kc, \kappa.x, \kappa.x_1)
		+ \varepsilon H(k, \kappa.x_1, \kappa.x_1 - \ell \kappa.k/\omega)
		\,,
\end{equation}
where the function $H$ is defined in \cref{eq:scalar propagator}.
For the scalar amplitude $\check \Aa = 1 + \varepsilon \check \Aa^\o1$, one obtains the transport equation
\begin{equation}
	\t \kc{^\mu} \t\p{_\mu} \check\Aa^\o1
		= - \half (\wop \check\psi)^\o1\,.
\end{equation}
A direct calculation shows that
\begin{equation}
	(\wop \check \psi)^\o1
		= \omegaG \omega (m.n) h(m,m)_{\kappa.x_1}
		- \half \omegaG \omega (1 + m.n) h(m,m)_{\kappa.x_1 - \ell \kappa.k/\omega}\,,
\end{equation}
so that the perturbation of the reflected amplitude is found to be
\begin{equation}
\begin{split}
	\check \Aa^\o1
		={}& -\half (m.n) \omegaG (\ell - m.x) h(m,m)_{\kappa.x_1}
		\\
		&+ \quarter (1 + m.n) \omegaG (2 \ell - m.x) h(m,m)_{\kappa.x_1 - \ell \kappa.k/\omega}
		\\&
		+ \delta \Aa(\tau.x_1 - \ell \tau.k/\omega, \xi.x_1 - \ell \xi.k/\omega, \zeta.x_1 - \ell \zeta.k/\omega)\,.
\end{split}
\end{equation}
To compute the reflected polarisation, we write
\begin{equation}
	\check f = (\dd \check \psi / \omega) \wedge \check E\,,
\end{equation}
in complete analogy to \cref{eq:polarisation f = k E}.
Since, according to \cref{eq:reflected wave vector}, the reflected eikonal satisfies
\begin{equation}
	\t \nabla{_\mu} \check \psi
		= \t R{_\mu^\rho} \t \nabla{_\rho} \psi
	\qquad(\text{on }\tilde\Sigma)\,,
\end{equation}
the last condition of \cref{eq:reflection conditions} can be solved by imposing
\begin{equation}
	\t{\check E}{_\mu}
		= - \t R{_\mu^\nu} \t{E}{_\nu}
	\qquad(\text{on }\tilde\Sigma)\,.
\end{equation}
%
The solution to the transport equation
\begin{equation}
	\t \kc{^\sigma} \t \p{_\sigma} \t {\check E}{^{\o1}_\mu}
		= - \t \Gamma{^{\o1}^\nu_\mu_\rho} \t \kc{^\rho} \t {\check E}{^{\o0}_\nu} \,,
\end{equation}
with this boundary value is then found to be given by
\begin{equation}
\begin{split}
	\t{\check E}{_\mu}
		={}&
		- \t E{^{\o0}_\mu}
		- \varepsilon \t{\varGamma(\kc, \kappa.x, \kappa.x_1)}{^\nu_\mu} \t E{^{\o0}_\nu}
		\\&
		- \varepsilon \t R{^{\o0}_\mu^\nu}\, \t{\varGamma(k, \kappa.x_1, \kappa.x_1- \ell \kappa.k/\omega)}{^\rho_\nu} \t E{^{\o0}_\rho}
		\\&
		- \varepsilon \t m{_\mu} \left(2 \t h{^j^k}(\kappa.x_1) - \t h{^j^k}(\kappa.x_1 - \ell \kappa.k/\omega) \right) \t m{_j} \t E{^{\o0}_k}
		\\&
		- \varepsilon \t {\delta E}{^\parallel_\mu}(\tau.x_1 - \ell \tau.k/\omega, \xi.x_1 - \ell \xi.k/\omega, \zeta.x_1 - \ell \zeta.k/\omega)\,,
\end{split}
\end{equation}
where, in the last two lines, we have used the decomposition of the emitted polarisation into normal and tangential parts, given in \cref{eq:polarisation perturbation decomposition}.

Finally, we evaluate the field back at the emission surface $\Sigma$, for which we introduce some notation. Given any point $x$ on $\Sigma$, denote by $x^R$ the point obtained by following the unperturbed incoming light ray back in time (along $- \kc$) until one reaches the mirror surface. Following the unperturbed light ray further back in time (along $-k$), one reaches the point $x^E$ on $\Sigma$, where the ray was initially emitted a time $2 \ell$ ago. Explicitly, the coordinates of these points are given by
\begin{align}
	\label{eq:points reflection emission}
	\t{(x^R)}{^\mu}
		&= \t x{^\mu} - \ell \t\kc{^\mu}/\omega\,,
	&
	\t{(x^E)}{^\mu}
		&= (t - 2  \ell,  \t x{^i})\,.
\end{align}
With this notation, the returning field on $\Sigma$ can be written as
\begin{align}
	\label{eq:returning field general}
	\check F|_\Sigma
		&= \check \Aa e^{i \check \psi} \, \check K \wedge \check E\,,
	\\
\shortintertext{where}
	\label{eq:returning field general eikonal}
	\check \psi
		&= 2 \omega \ell - \omega t
		+ \varepsilon H(\kc, \kappa.x, \kappa.x^R)
		+ \varepsilon H(k, \kappa.x^R, \kappa.x^E)\,,
	\\
	\begin{split}
		\check \Aa
			&=
			1 - \half \varepsilon \omegaG \ell [
				(m.n) h'(m,m)_{\kappa.x^R}
				- (1 + m.n)  h'(m,m)_{\kappa.x^E}
			]
			+ \varepsilon \delta \Aa(x^E)
			\,,
	\end{split}
	\\
	\label{eq:returning field general wave vector}
	\begin{split}
		\check K
			&= - \dd t - \nu
			+ \half \varepsilon
			\left[
				\omega \kappa + m (\kappa.\kc)
			\right]
			\left[
				\frac{h(m,m)}{\kappa.\kc} \bigg|_{\kappa.x^R}^{\kappa.x}
				+ \frac{h(m,m)}{\kappa.k} \bigg|^{\kappa.x^R}_{\kappa.x^E}
			\right]
			\,,
	\end{split}
	\\
	\label{eq:returning field general polarisation}
	\begin{split}
		\check E
			&=
			- \left(
				\mathbf 1
				+ \varepsilon \varGamma(\kc, \kappa.x, \kappa.x^R)
				+ \varepsilon R^\o0 \varGamma(k, \kappa.x^R, \kappa.x^E)
			\right)
			\left(
				E^\o0
				+ \varepsilon \delta E^\parallel(x^E)
			\right)
			\\&\qquad
			- \varepsilon m \left(
				2 h(m, E^\o0)_{\kappa.x^R}
				- h(m,E^\o0)_{\kappa.x^E}
			\right)
			\,,
	\end{split}
\end{align}
where the subscripts indicate the values of the parameter $u$ in the metric perturbation $h$, and the last expression is understood in matrix notation (i.e.\ $\mathbf 1$ is the identity matrix, $\varGamma$ is the matrix defined in \cref{eq:tensor propagator}, and $R^\o0 \varGamma$ denotes the matrix product of $R^\o0$ and $\varGamma$ as defined in \cref{eq:reflection tensor,eq:tensor propagator}).

\section{Non-Polarising Beam Splitters}
\label{s:michelson beam splitter}

Considering now a non-polarising 50:50 beam splitter, we model the two “output ports” by timelike hypersurfaces $\Sigma_\cI$ and $\Sigma_\cII$ (as sketched in \cref{fig:michelson}), whose unperturbed normals $\mI$ and $\mII$ are orthogonal in the background geometry.
The associated unit one-forms (in the perturbed geometry) are
\begin{align}
	\t \nuI{_i}
	&= \t m{_i} \left( 1 + \half \varepsilon h(\mI,\mI) \right)\,,
	&
	\t \nuII{_i}
	&= \t \mII{_i} \left( 1 + \half \varepsilon h(\mII,\mII) \right)\,,
\end{align}
c.f.\ \cref{eq:normal perturbed}. 
Radiation sent into the beam splitter partially passes through, and is partially deflected. This deflection can be described using the same methods as discussed in the previous section, namely by reflecting the incoming field strength tensor along the normal
\begin{equation}
	\tilde \nu = (\nuII - \nuI)/\|\nuII - \nuI\|\,,
\end{equation}
cf.\ the description of reflection in \cref{appendix:perfect reflection}.
To this end, define the reflection operator
\begin{equation}
	\t{{\Rbs}}{_i^j}
		= \t* \delta{_i^j} - 2 \t {\tilde \nu}{_i} \t {\tilde \nu}{^j}
		= \t* \delta{_i^j}
			- \frac{1}{1 - g(\nuI, \nuII)}
			\t{(\nuII-\nuI)}{_i} \t{(\nuII-\nuI)}{_k} \t g{^k^j}
	\,,
\end{equation}
which acts by interchanging $\nuI$ and $\nuII$, and leaving all vectors orthogonal to them unchanged.
In flat space, $\t{{\Rbs}}{_i^j}$ reduces to
\begin{equation}
	\t{{\Rbs}}{^{\o0}_i^j}
		= \t* \delta{_i^j} - \t{(\mII - \mI)}{_i} \t{(\mII - \mI)}{^j}
	\,.
\end{equation}

With this, we can now formulate a simple model of the beam splitter.
Consider incoming radiation at the surface $\Sigma_{\textsc{in}}$ (coming from a laser) of the form
\begin{equation}
	\label{eq:laser field}
	F_\textsc{in}
		= \sqrt 2 K_\textsc{in} \wedge E_\textsc{in} e^{i \psi}\,,
\end{equation}
where $E_\textsc{in}$ is as in \cref{eq:emitted polarisation} and the normalisation is such that the emitted radiation has unit energy density.

We then assume the fields at the output ports (facing the mirrors) to be
\begin{align}
	\label{eq:beam splitting}
	\text{on }\Sigma_\cI:&\quad
	\tfrac{1}{\sqrt 2} K_\cI \wedge E_\cI e^{i \psi}\,,
	&
	\text{on }\Sigma_\cII:&\quad
	\tfrac{1}{\sqrt 2} K_\cII \wedge E_\cII e^{i \psi}\,.
\end{align}
The outgoing fields have only half the amplitude of the incoming field, in accordance with energy conservation.
This means that half the radiation passes through the beam splitter unchanged, while the other half is reflected along the normal $\tilde \nu$ defined above.

Here, $K_{\cI,\cII} = \omega(\nu_{\cI,\cII} - \dd t)$ are the wave vectors normal to the respective surfaces, and the polarisation vectors are solutions of the respective constraint equations which satisfy the following relations \emph{at the spatial origin}
\begin{align}
	E_\cI &= E_\textsc{in}\,,
	&
	E_\cII &= - \Rbs E_\textsc{in}\,.
\end{align}
To obtain an expression for the amplitude everywhere on $\Sigma$, one must then construct a solution to the constraint equations with this prescribed behaviour at the spatial origin. The method of doing so is described in \Cref*{appendix:constraints}, but the precise details are irrelevant for the applications where.

These fields propagate towards the mirrors, are reflected there, and return to the beam splitter.
In the course of this propagation, the fields undergo changes as described by the equations \eqref{eq:returning field general eikonal}—\eqref{eq:returning field general polarisation}.
Irrespective of the precise form of this transformation, the returning radiation (restricted to $\Sigma_\cI$ and $\Sigma_\cII$) can be written in the form
\begin{align}
	\text{on }\Sigma_\cI:&\quad
	\tfrac{1}{\sqrt 2} \check K_\cI \wedge \check E_\cI e^{i \check \psi_\cI}\,,
	&
	\text{on }\Sigma_\cII:&\quad
	\tfrac{1}{\sqrt 2} \check K_\cII \wedge \check E_\cII e^{i \check \psi_\cII}\,,
\end{align}
Finally, these fields pass through the beam splitter to reach the output port. To model this, we assume the radiation impinging at $\Sigma_\cI$ to be deflected as above, while the radiation impinging at $\Sigma_\cII$ is inverted because of the $\pi$ phase shift at a beam splitter.
This translates to the following formula for the output field at the spatial origin:
\begin{equation}
	\label{eq:beam merging}
	F_{\textsc{out}}
		= - \tfrac{1}{\sqrt 2} (\Rbs \check K_\cI) \wedge (\Rbs \check E_\cI) e^{i \check \psi_\cI}
		- \tfrac{1}{\sqrt 2} \check K_\cII \wedge \check E_\cII e^{i \check \psi_\cII}\,.
\end{equation}
%

\section{Interferometer Output Signal}
\label{s:michelson output}

Having developed a description of light propagation, reflection at mirrors and at beam splitters, all in the presence of a plane gravitational wave, we can now compute the output of a simple Michelson interferometer as sketched in \cref{fig:michelson}.

\paragraph{Ray I}

One ray is first transmitted by the beam splitter, then reflects at the mirror surface ${\{ \mI.x = \ell_\cI \}}$, and propagates back to the beam splitter where it is finally deflected towards the detector.

The radiation sent into the interferometer arm is precisely of the form \eqref{eq:emission bdry eikonal}—\eqref{eq:emission bdry polarisation}, rescaled only by a factor $1/\sqrt{2}$ in accordance with \cref{eq:beam splitting}. Up to this global factor, the field returning to the beam splitter is thus obtained from \cref{eq:returning field general} by the substitution
\begin{align}
	m &\to m_\cI\,,
	&
	\ell &\to \ell_\cI\,,
	&
	k &\to k_\cI\,,
	&
	x^E &\to x^E_\cI\,,
	&
	x^R &\to x^R_\cI\,,
\end{align}
where $x^E_\cI$ and $x^R_\cI$ are defined as in \cref{eq:points reflection emission}, except for this same substitution:
\begin{align}
	\t{(x^R_\cI)}{^\mu}
		&= \t x{^\mu} - \ell_\cI \t*\kc{_\cI^\mu}/\omega\,,
	&
	\t{(x^E_\cI)}{^\mu}
		&= (t - 2  \ell_\cI, \t x{^i})\,.
\end{align}
According to \cref{eq:beam merging}, the corresponding radiation reaching the output port of the beam splitter is obtained by applying $\Rbs$ to both the local wave vector and the polarisation, and changing the overall sign of the latter.
A direct calculation shows that
\begin{equation}
	\label{eq:beam splitter intermediate}
	\Rbs^\o1 E^\o0
		= \half \mI h(\mI - \mII, E^\o0 + \Rbs^\o0 E^\o0)
		+ \mII [ h(\mII, \Rbs^\o0 E^\o0) - h(\mI, E^\o0)]\,,
\end{equation}
so that the contribution of this first ray to the output field (omitting the háček) is
\begin{align}
	\label{eq:michelson beam 1}
	F_\cI
		&= \Aa_\cI\, K_\cI \wedge E_\cI\, e^{i \psi_\cI} \,,
\shortintertext{where}
	\label{eq:michelson S1}
	\psi_\cI
		&= 2 \omega \ell_\cI - \omega t
		+ \varepsilon H(\kc_\cI, \kappa.x, \kappa.x^R_\cI)
		+ \varepsilon H(k_\cI, \kappa.x^R_\cI, \kappa.x^E_\cI)
		\,,
	\\
	\label{eq:michelson A1}
	\Aa_\cI
		&= \frac{1}{\sqrt 2} - \frac{\varepsilon \omegaG \ell_\cII}{2 \sqrt2} \left[
			(n.\mI)\, h'(\mI,\mI)_{\kappa.x^R_\cI}
			- (1+n.\mI) h'(\mI,\mI)_{\kappa.x^E_\cI}
		\right]
		+ \varepsilon \delta \Aa(x^E_\cI)\,,
	\\
	\label{eq:michelson K1}
	\begin{split}
		K_\cI
			&= - \dd t - \nuII
			+ \half \varepsilon
			\left[
				\omega \Rbs^\o0 \kappa
				+ \mII (\kappa.\kc_\cI)
			\right]
			\left[
				\frac{h(\mI,\mI)}{\kappa.\kc_\cI} \bigg|_{\kappa.x^R_\cI}^{\kappa.x}
				+ \frac{h(\mI,\mI)}{\kappa.k_\cI} \bigg|^{\kappa.x^R_\cI}_{\kappa.x^E_\cI}
			\right]\,,
	\end{split}
	\\
	\label{eq:michelson E1}
	\begin{split}
		E_\cI
		&=
		\Rbs^\o0 \left(
			\mathbf 1
			+ \varepsilon \varGamma(\kc_\cI, \kappa.x, \kappa.x^R_\cI)
			+ \varepsilon \RI^\o0 \varGamma(k_\cI, \kappa.x^R_\cI, \kappa.x^E_\cI)
		\right) E^\o0
		+ \varepsilon \Rbs^\o0 \delta E^\parallel(x^E_\cI)
		\\&
		- \half \varepsilon \mI\, h(\mII - \mI, E^\o0 + \Rbs^\o0 E^\o0)_{\kappa.x}
		\\&
		+ \varepsilon \mII \left(
			2 h(\mI, E^\o0)_{\kappa.x^R_\cI}
			- h(\mI,E^\o0)_{\kappa.x^E_\cI}
			- h(\mI,E^\o0)_{\kappa.x}
			+ h(\mII, \Rbs^\o0 E^\o0)_{\kappa.x}
		\right) \,.
	\end{split}
\end{align}

\paragraph{Ray II}
The second ray is first deflected by the beam splitter, reflects at the mirror surface ${\{\mII.x = \ell_\cII \}}$ and is then transmitted by the beam splitter.
In analogy to the first ray, we consider the expression obtained from \cref{eq:returning field general} by the substitution
\begin{align}
	m &\to m_\cII\,,
	&
	\ell &\to \ell_\cII\,,
	&
	k &\to k_\cII\,,
	&
	x^E &\to x^E_\cII\,,
	&
	x^R &\to x^R_\cII\,,
\end{align}
where
\begin{align}
	\t{(x^R_\cII)}{^\mu}
		&= \t x{^\mu} - \ell_\cII \t*\kc{_\cII^\mu}/\omega\,,
	&
	\t{(x^E_\cII)}{^\mu}
		&= (t - 2  \ell_\cII, \t x{^i})\,.
\end{align}
However, due to the deflection at the beam splitter \emph{before} the emission in the detector arm, we must also substitute
\begin{align}
	E^\o0
		&\to - \Rbs^\o0 E^\o0\,,
	\\
	\delta E^\parallel
		&\to - \Rbs^\o0 \delta E^\parallel
		+ \half \mI h(\mII - \mI, E^\o0 + \Rbs^\o0 E^\o0)
	\,,
\end{align}
as follows from \cref{eq:beam splitter intermediate}. This leads to the following expression for the contribution of the second ray to the interferometer output:
\begin{align}
	\label{eq:michelson beam 2}
	F_\cII
		&= \Aa_\cII\, K_\cII \wedge E_\cII\, e^{i \psi_\cII} \,,
\shortintertext{where}
	\label{eq:michelson S2}
	\psi_\cII
		&= 2 \omega \ell_\cII - \omega t
		+ \pi
		+ \varepsilon H(\kc_\cII, \kappa.x, \kappa.x^R_\cII)
		+ \varepsilon H(k_\cII, \kappa.x^R_\cII, \kappa.x^E_\cII)
	\,,
	\\
	\label{eq:michelson A2}
	\Aa_\cII
		&= \frac{1}{\sqrt 2} - \frac{\varepsilon \omegaG \ell_\cII}{2 \sqrt2} \left[
			(n.\mII) h'(\mII,\mII)_{\kappa.x^R_\cII}
			- (1+n.\mII) h'(\mII,\mII)_{\kappa.x^E_\cII}
		\right]
		+ \varepsilon \delta \Aa(x^E_\cII)\,,
	\\
	\label{eq:michelson K2}
	\begin{split}
		K_\cII
			&= - \dd t - \nuII
			+ \half \varepsilon
			\left[
				\omega \kappa
				+ \mII (\kappa.\kc_\cII)
			\right]
			\left[
				\frac{h(\mII,\mII)}{\kappa.\kc_\cII} \bigg|_{\kappa.x^R_\cII}^{\kappa.x}
				+ \frac{h(\mII,\mII)}{\kappa.k_\cII} \bigg|^{\kappa.x^R_\cII}_{\kappa.x^E_\cII}
			\right]
			\,,
	\end{split}
	\\
	\label{eq:michelson E2}
	\begin{split}
		E_\cII
		&=
		\left(
			\mathbf 1
			+ \varepsilon \varGamma(\kc_\cII,\kappa.x, \kappa.x^R_\cII)
			+ \varepsilon {\RII}^\o0 \varGamma(k_\cII,\kappa.x^R_\cII, \kappa.x^E_\cII)
		\right) \Rbs^\o0 E^\o0
		+ \varepsilon \Rbs^\o0 \delta E^\parallel(x^E_\cII)
		\\&
		- \half \varepsilon \mI\, h(\mII - \mI, E^\o0 + \Rbs^\o0 E^\o0)_{\kappa.x^E_\cII}
		\\&
		+ \varepsilon \mII \left(
			2 h(\mII, \Rbs^\o0 E^\o0)_{\kappa.x^R_\cII}
			- h(\mII, \Rbs^\o0 E^\o0)_{\kappa.x^E_\cII}
		\right)
		\,.
	\end{split}
\end{align}
Here, we have decided to put the $\pi$ phase shift at the beam splitter into the eikonal so that all amplitudes are positive.

These results can now be used to compute the final output field.

\paragraph{Interferometer Output}

So far, we have worked with a complex electromagnetic field for simplicity. The physical field, however, is obtained by taking the real part. Assuming the emitted polarisation $E$ to be real (corresponding to linear polarisation), the real fields are obtained simply by replacing $e^{i \psi}$ by $\cos(\psi)$ in \cref{eq:michelson beam 1,eq:michelson beam 2}.
The \emph{overall} real field at the output port is thus
\begin{equation}
	\label{eq:detector field}
	\Ff
		= \Aa_\cI\, K_\cI \wedge E_\cI\, \cos(\psi_\cI)
		+ \Aa_\cII\, K_\cII \wedge E_\cII\, \cos(\psi_\cII)\,.
\end{equation}

There are multiple readout schemes to measure the \GW\ perturbations of this resulting field, notably the homodyne, heterodyne, and \textsc{dc} readout schemes \cites{Fritschel2001,Fricke2012,Fritschel2014}.
While the homodyne readout scheme was used in the first generation of \GW\ detectors, the  Advanced~\textsc{ligo}, Advanced~Virgo, \textsc{kagra} and \textsc{geo 600} detectors now use the \textsc{dc} readout scheme (\DCR) \cites{Buikema2020,Acernese2014,Akutsu2020,Degallaix2010}. However, as the planned \textsc{ligo a+} upgrade will transition to the balanced homodyne readout scheme (\BHR) \cite{Fritschel2020}, we will describe both the \DCR\ and \BHR\ schemes in the following analysis.

In all cases, the signal is read out using photo diodes, which are sensitive to the energy density of the \EM\ field.
As shown in \cref{appendix:output power}, the time-averaged intensity of a field of the form \eqref{eq:detector field}
at first order in $\varepsilon$ is
\begin{equation}
	\label{eq:field energy 2 rays}
	\braket{\t T{_0_0}}
		= \half (\Aa_\cI\, \t K{_\cI_0})^2
		+ \half (\Aa_\cII\, \t K{_\cII_0})^2
		+ (\Aa_\cI \t K{_\cI_0}) (\Aa_\cII \t K{_\cII_0})  \cos(\psi_\cI - \psi_\cII)\,,
\end{equation}
where, notably, no inner products of the wave vectors and/or polarisation vectors enter. This is because linear perturbations of these normalised vectors produce only quadratic corrections to the relevant inner products and are thus negligible.
This means that although the gravitational wave causes perturbations of the eikonal, amplitude, local wave vector, and the polarisation, only the first three perturbations produce a signal in the detectors considered.

Note that we have chosen the input field to be normalised to unit energy density. In the general case, the expression given here is the output power normalised to the input power.

\Cref{eq:field energy 2 rays}, as well as its generalisation to more than two interfering rays, can be used to compute the signals in the \DCR\ and \BHR\ schemes. Here and in the following, all expressions are evaluated at the spatial coordinate origin, where the detector is positioned.

\paragraph{DC Readout Scheme}

In the \DCR\ scheme, one directly measures the intensity of the interferometer output, yielding the signal
\begin{equation}
	S_\DCR
		= \half (\Aa_\cI\, \t K{_\cI_0})^2
		+ \half (\Aa_\cII\, \t K{_\cII_0})^2
		+ (\Aa_\cI \t K{_\cI_0}) (\Aa_\cII \t K{_\cII_0})  \cos(\psi_\cI - \psi_\cII)\,.
\end{equation}
In the unperturbed case, this reduces to
\begin{equation}
	S_\DCR^\o0
		= \half - \half \cos(2 \omega(\ell_\cI - \ell_\cII))
		= \sin^2(\omega(\ell_\cI - \ell_\cII))
		\,,
\end{equation}
which vanishes if both arm lengths are chosen to be equal.
The gravitational wave produces three correction terms, corresponding to perturbations of the eikonal, amplitude, and wave vector:
\begin{equation}
	S_\DCR
		= S_\DCR^\o0
		+ \varepsilon S_{\DCR, \psi}
		+ \varepsilon S_{\DCR, \Aa}
		+ \varepsilon S_{\DCR, K}
		+ O(\varepsilon^2)\,.	
\end{equation}
Using \crefrange{eq:michelson S1}{eq:michelson K1},  \crefrange*{eq:michelson S2}{eq:michelson K2}, as well as \cref{eq:amplitude correction origin}, these first-order perturbations evaluate to
\begin{align}
	\label{eq:signal DC phase full}
	\begin{split}
		S_{\DCR, \psi}
			={}& + \half \sin(2 \omega(\ell_\cI - \ell_\cII))
			\\&\times
			\bigg(
				H(\kc_\cI, \kappa.x, \kappa.x^R_\cI)
				+ H(k_\cI, \kappa.x^R_\cI, \kappa.x^E_\cI)
				\\&\qquad
				- H(\kc_\cII, \kappa.x, \kappa.x^R_\cII)
				- H(k_\cII, \kappa.x^R_\cII, \kappa.x^E_\cII)
			\bigg)\,,
	\end{split}\\
	\label{eq:signal DC amplitude full}
	\begin{split}
		S_{\DCR, \Aa}
			={}& - \half \sin^2(\omega(\ell_\cI - \ell_\cII))
			\\&\times
			\bigg(
				\omega \ell_\cI \left[
					(n.\mI)\, h'(\mI,\mI)_{\kappa.x^R_\cI}
					- (1+n.\mI) h'(\mI,\mI)_{\kappa.x^E_\cI}
				\right]
				\\&\qquad
				+ \omega \ell_\cII \left[
					(n.\mII)\, h'(\mII,\mII)_{\kappa.x^R_\cII}
					- (1+n.\mII) h'(\mII,\mII)_{\kappa.x^E_\cII}
				\right]
			\bigg)\,,
	\end{split}\\
	\label{eq:signal DC frequency full}
	\begin{split}
		S_{\DCR, K}
			={}& - \half \omegaG \omega \sin^2(\omega(\ell_\cI - \ell_\cII))
			\\&\times
			\bigg(
				\frac{h(\mI,\mI)}{\kappa.\kc_\cI} \bigg|_{\kappa.x^R_\cI}^{\kappa.x}
				+ \frac{h(\mI,\mI)}{\kappa.k_\cI} \bigg|^{\kappa.x^R_\cI}_{\kappa.x^E_\cI}
				+ \frac{h(\mII,\mII)}{\kappa.\kc_\cII} \bigg|_{\kappa.x^R_\cII}^{\kappa.x}
				+ \frac{h(\mII,\mII)}{\kappa.k_\cII} \bigg|^{\kappa.x^R_\cII}_{\kappa.x^E_\cII}
			\bigg)\,,
	\end{split}
\end{align}
all of which vanish for equal arm lengths. Hence, the \DCR\ scheme requires a slight asymmetry $\Delta \ell = \ell_\cI - \ell_\cII$ to obtain a linear response to the \GW\ perturbation.

The three terms appearing here can be interpreted as follows.
$S_{\DCR, \psi}$ arises due to the perturbation of the optical path lengths along the two interferometer arms. The multiplicative factor ${\sin(2 \omega( \ell_\cI - \ell_\cII))}$ entails that the phase response vanishes when the unperturbed output power is extremal, and is maximal in between.
The second term, $S_{\DCR, \Aa}$, arises from amplitude perturbations, which can be regarded as scintillations of the light signal, which grow with the length of the interferometer arms (for very long distances, corrections from higher order terms of the geometrical optics expansion might become relevant). The strength of this effect is directly proportional to the output intensity in the unperturbed configuration, as can be seen from the multiplicative factor $\sin^2(\omega(\ell_\cI - \ell_\cII))$. Finally, the last term, $S_{\DCR, K}$, arises because the gravitational wave modulates the frequency of the light rays, which translates to a modulation of the output power, similar to the amplitude modulation.

Hence, the first contribution, $S_{\DCR, \psi}$, describes the standard phase response (as it is determined from the eikonal alone), while the remaining terms describe the corrections beyond the eikonal equation.

\paragraph{Balanced Homodyne Readout Scheme}

In the \BHR\ scheme, one mixes the output field \eqref{eq:detector field} with a further ray — typically derived from the same laser source — which is not sent through the interferometer.
For this ray the \GW\ perturbations are negligible so that one may set
\begin{align}
	\psi_\BHR &= - \omega t + \varphi
	&
	\Aa_\BHR &= \sqrt{2}\,,
	&
	K_\BHR &= - \dd t + \nu_\cI\,,
\end{align}
where $\varphi$ denotes the readout angle (an experimentally controlled parameter), and the amplitude normalisation is the same as in \cref{eq:laser field}.
Sending this ray and the output field \eqref{eq:detector field} orthogonally through a  further 50:50 beam splitter, the resulting intensities at the two output ports evaluate to
\begin{align}
\begin{split}
	I
		={}& \quarter S_\DCR
		+ \eighth (\Aa_\BHR\, \t K{_\BHR_0})^2
		\\&
		+ \quarter (\Aa_\BHR \t K{_\BHR_0}) \left[
			(\Aa_\cI \t K{_\cI_0})  \cos(\psi_\BHR - \psi_\cI)
			+ (\Aa_\cII \t K{_\cII_0})  \cos(\psi_\BHR - \psi_\cII)
		\right]
	\,,
\end{split}\\
\begin{split}
	I'
		={}& \quarter S_\DCR
		+ \eighth (\Aa_\BHR\, \t K{_\BHR_0})^2
		\\&
		- \quarter (\Aa_\BHR \t K{_\BHR_0}) \left[
			(\Aa_\cI \t K{_\cI_0})  \cos(\psi_\BHR - \psi_\cI)
			+ (\Aa_\cII \t K{_\cII_0})  \cos(\psi_\BHR - \psi_\cII)
		\right]
	\,,
\end{split}
\end{align}
where the difference in sign of the interference terms involving the third ray arises because of the $\pi$ phase shift at the beam splitter.
The signal in the \BHR\ scheme is then obtained by measuring the difference of the two intensities, i.e.\
\begin{equation}
	S_\BHR
		=
		\half (\Aa_\BHR \t K{_\BHR_0}) \left(
			\Aa_\cI \t K{_\cI_0} \cos(\psi_\BHR - \psi_\cI)
			+ \Aa_\cII \t K{_\cII_0} \cos(\psi_\BHR - \psi_\cII)
		\right)\,.
\end{equation}
The unperturbed signal in this readout scheme is
\begin{equation}
	S_\BHR^\o0
		= \half\left[
			\cos(\varphi - 2 \omega \ell_\cI)
			- \cos(\varphi - 2 \omega \ell_\cII)
		\right]\,,
\end{equation}
which vanishes for all values of $\varphi$ if the arm lengths are chosen to be equal.

As for the \DCR\ scheme, it is convenient to separate the first-order perturbations of the phase, amplitude, and frequency by writing
\begin{equation}
	S_\BHR
		= S_\BHR^\o0
		+ \varepsilon S_{\BHR, \psi}
		+ \varepsilon S_{\BHR, \Aa}
		+ \varepsilon S_{\BHR, K}
		+ O(\varepsilon^2)\,,
\end{equation}
which, for equal arm lengths, evaluate to
\begin{align}
	\label{eq:signal BHR phase full}
	\begin{split}
		S_{\BHR, \psi}
			={}& + \half \sin(\varphi - 2 \omega \ell)
			\\&\times
			\bigg(
				H(\kc_\cI, \kappa.x, \kappa.x^R_\cI)
				+ H(k_\cI, \kappa.x^R_\cI, \kappa.x^E_\cI)
				\\&\qquad
				- H(\kc_\cII, \kappa.x, \kappa.x^R_\cII)
				- H(k_\cII, \kappa.x^R_\cII, \kappa.x^E_\cII)
			\bigg)\,,
	\end{split}\\
	\label{eq:signal BHR amplitude full}
	\begin{split}
		S_{\BHR, \Aa}
			={}& - \quarter \omegaG \ell \cos(\varphi - 2 \omega \ell)
			\\&\times
			\bigg(
				(n.\mI)\, h'(\mI,\mI)_{\kappa.x^R_\cI}
				- (1+n.\mI) h'(\mI,\mI)_{\kappa.x^E_\cI}
				\\&\qquad
				-  (n.\mII)\, h'(\mII,\mII)_{\kappa.x^R_\cII}
				+ (1+n.\mII) h'(\mII,\mII)_{\kappa.x^E_\cII}
			\bigg)\,,
	\end{split}\\
	\label{eq:signal BHR frequency full}
	\begin{split}
		S_{\BHR, K}
			={}& - \quarter \omegaG \omega \cos(\varphi - 2 \omega \ell)
			\\&\times
			\bigg(
				\frac{h(\mI,\mI)}{\kappa.\kc_\cI} \bigg|_{\kappa.x^R_\cI}^{\kappa.x}
				+ \frac{h(\mI,\mI)}{\kappa.k_\cI} \bigg|^{\kappa.x^R_\cI}_{\kappa.x^E_\cI}
				- \frac{h(\mII,\mII)}{\kappa.\kc_\cII} \bigg|_{\kappa.x^R_\cII}^{\kappa.x}
				- \frac{h(\mII,\mII)}{\kappa.k_\cII} \bigg|^{\kappa.x^R_\cII}_{\kappa.x^E_\cII}
			\bigg)\,,
	\end{split}
\end{align}
While the phase response in the \BHR\ scheme \eqref{eq:signal BHR phase full} is the same as in the \DCR\ scheme \eqref{eq:signal DC phase full}, the amplitude and frequency responses differ. Specifically, \cref{eq:signal DC amplitude full,eq:signal DC frequency full} contain the sum of the perturbations in the two interferometer arms, while \cref{eq:signal BHR amplitude full,eq:signal BHR frequency full} contain their difference.

\section{Detector Pattern Functions in the Low Frequency Limit}
\label{s:pattern functions}

In the low frequency approximation, which applies if the \GW\ wavelength is much longer than the interferometer arms, the above formulae can be simplified significantly by assuming the \GW\ waveform to vary slowly along the light ray trajectories. Formally, this corresponds to a Taylor expansion of $h$ near $x = 0$ to leading order (i.e.\ including the first non-trivial term).

To set up the notation, we first consider the well-known phase perturbations in \cref{eq:signal DC phase full,eq:signal BHR phase full}. To analyse the phase response of the interferometer in both readout schemes, set
\begin{equation}
	S_\psi
		= \half \left[
		H(\kc_\cI, \kappa.x, \kappa.x^R_\cI)
		+ H(k_\cI, \kappa.x^R_\cI, \kappa.x^E_\cI)
		- H(\kc_\cII, \kappa.x, \kappa.x^R_\cII)
		- H(k_\cII, \kappa.x^R_\cII, \kappa.x^E_\cII)
		\right]\,,
\end{equation}
so that
\begin{align}
	S_{\DCR,\psi}
		&= \sin(2 \omega \Delta \ell) S_\psi\,,
		&
	S_{\BHR,\psi}
		&= \sin(\varphi - 2 \omega \ell) S_\psi\,,
\end{align}
Assuming (almost) equal arm lengths, the low-frequency limit of $S_\psi$ evaluates to
\begin{equation}
	S_\psi \approx
		\half \omega \ell \left(
			h(\mI, \mI) - h(\mII, \mII)
		\right)\,,
\end{equation}
where the metric perturbation, $h$, is evaluated at the coordinate origin.
Decomposing $h$ in terms of the two polarisations “$+$” and “$\times$” (described in more detail below) as
\begin{equation}
	\label{eq:GW polarisation decomposition}
	\t h{_\mu_\nu}(u)
		= \t* A{^+_\mu_\nu}\, h_+(u)
		+ \t* A{^\times_\mu_\nu}\, h_\times(u)\,,
\end{equation}
one can write
\begin{align}
	\label{eq:response phase}
	S_\psi
		&\approx \omega \ell \left(
			F_\psi^+ h_+ + F_\psi^\times h_\times
		\right)\,,
\shortintertext{where}
	\label{eq:pattern fctn phase def}
	F_\psi^+
	&= \half [A^+(\mI, \mI) - A^+(\mII, \mII)]\,,
	&
	F_\psi^\times
		&= \half [A^\times(\mI, \mI) - A^\times(\mII, \mII)]\,.
\end{align}
The functions $F_\psi^+$ and $F_\psi^\times$ are known as the detector pattern functions).
To describe their behaviour for various incidence angles of the gravitational wave, one must parameterise the \GW\ wave vector as well as the \GW\ polarisation tensors $A^+$ and $A^\times$. Specifically, one must relate the detection coordinate system $xyz$ (where the interferometer arms span the $x$ and $y$ axes) to a suitable coordinate system $XYZ$ associated to the gravitational wave.
\begin{figure}[ht]
	\centering
	\includegraphics[]{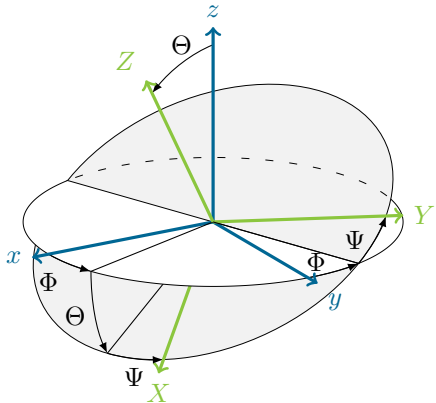}
	\caption{Illustration of the Euler angles $\Phi$, $\Theta$ and $\Psi$ in the “ZYZ” convention.}
	\label{fig:Euler angles}
\end{figure}
For this purpose, it is customary to use Euler angles of the kind “ZYZ”, as depicted in \cref{fig:Euler angles}. The \GW\ coordinate axes $XYZ$ are then obtained from the detector axes $xyz$ by first rotating by an angle $\Phi$ about the $z$-axis, then by an angle $\Theta$ about the transformed $y$ axis, and finally by an angle $\Psi$ about the such transformed $z$ axis.
The associated matrix which transforms the \GW\ coordinates to the detector coordinates is
\begin{equation}
	R(\Phi, \Theta, \Psi)
		=
		\begin{pmatrix}
			\cos \Phi	&	- \sin \Phi				&	0\\
			\sin \Phi	& 	\phantom{-}  \cos \Phi	&	0\\
			0			&	\phantom{-}0			&	1
		\end{pmatrix}
		\begin{pmatrix}
			\phantom{-} \cos\Theta	&	0	&	\sin\Theta\\
			0						&	1	&	0\\
			- \sin\Theta			&	0	&	\cos\Theta
		\end{pmatrix}
		\begin{pmatrix}
			\cos \Psi	&	- \sin \Psi				&	0\\
			\sin \Psi	& 	\phantom{-}  \cos \Psi	&	0\\
			0			&	\phantom{-}0			&	1
		\end{pmatrix}\,.
\end{equation}
For comparison, we note that the transformation matrix given in Ref.~\cite{Schutz1987} uses the “ZXZ” convention for Euler angles and is thus obtained from the one here by the substitution ${\Phi \to \Phi - \pi/2}$,  ${\Psi \to \Psi + \pi/2}$.
To parameterise the \GW\ polarisation tensors, denote by $\ee_1, \ee_2, \ee_3$ the canonical orthonormal basis of $\rNumbers^3$.
The $X$ and $Y$ basis vectors of the \GW\ frame then correspond to the following vectors in the detection frame:
\begin{align}
	\mathfrak b_1
		&= R(\Phi, \Theta, \Psi) \ee_1\,,
	&
	\mathfrak b_2
		&= R(\Phi, \Theta, \Psi) \ee_2\,,
\end{align}
using which one can define the polarisation tensors of \cref{eq:GW polarisation decomposition} as
\begin{align}
	\label{eq:GW polarisation parametrisation}
	A^+
		&= \mathfrak b_1 \otimes \mathfrak b_1 - \mathfrak b_2 \otimes \mathfrak b_2\,,
	&
	A^\times
		&= \mathfrak b_1 \otimes \mathfrak b_2 + \mathfrak b_2 \otimes \mathfrak b_1\,.
\end{align}
These quantities depend implicitly on all Euler angles $\Phi, \Theta, \Psi$, but it suffices to consider $\Psi = 0$ since the general case is obtained from the particular using the formula
\begin{equation}
	\begin{pmatrix}
		A^+(\Phi, \Theta, \Psi)\\
		A^\times(\Phi, \Theta, \Psi)
	\end{pmatrix}
	=
	\begin{pmatrix}
		\phantom{-}\cos(2 \Psi)	&	\sin(2 \Psi)\\
		- \sin(2\Psi)	& 	\cos(2 \Psi)
	\end{pmatrix}
	\begin{pmatrix}
		A^+(\Phi, \Theta, 0)\\
		A^\times(\Phi, \Theta, 0)
	\end{pmatrix}\,.
\end{equation}

\begin{figure}[ht]
	\centering
	\begin{subfigure}[c]{0.4\textwidth}
		\includegraphics[width=\columnwidth]{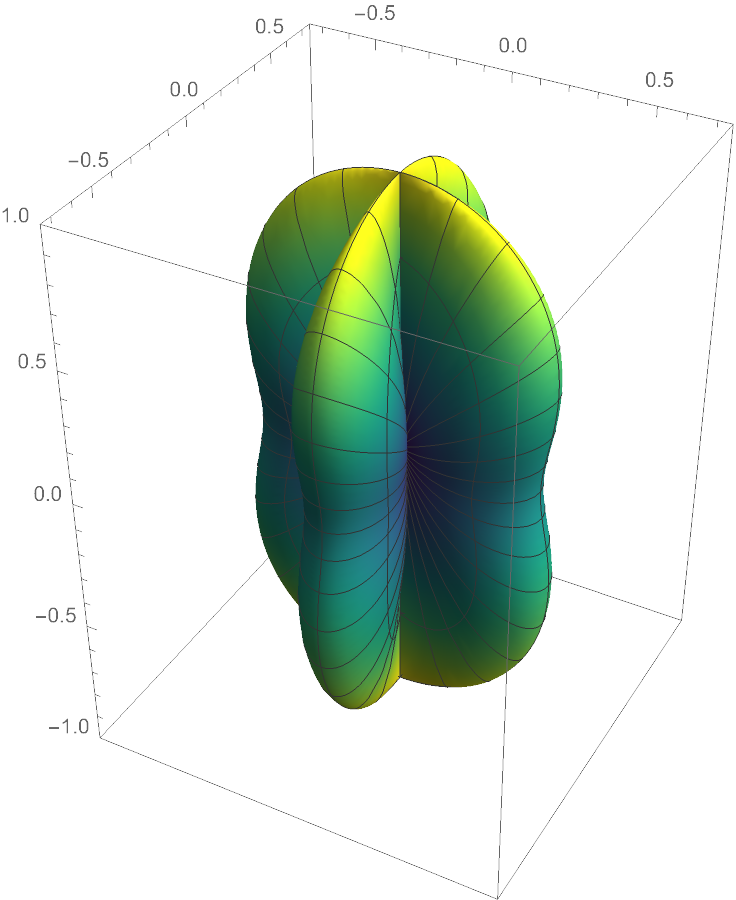}
		\subcaption{Eikonal pattern function $|F_\psi^+|$}
		\label{fig:pattern eikonal +}
	\end{subfigure}
	\hspace{0.1\textwidth}
	\begin{subfigure}[c]{0.4\textwidth}
		\includegraphics[width=\columnwidth]{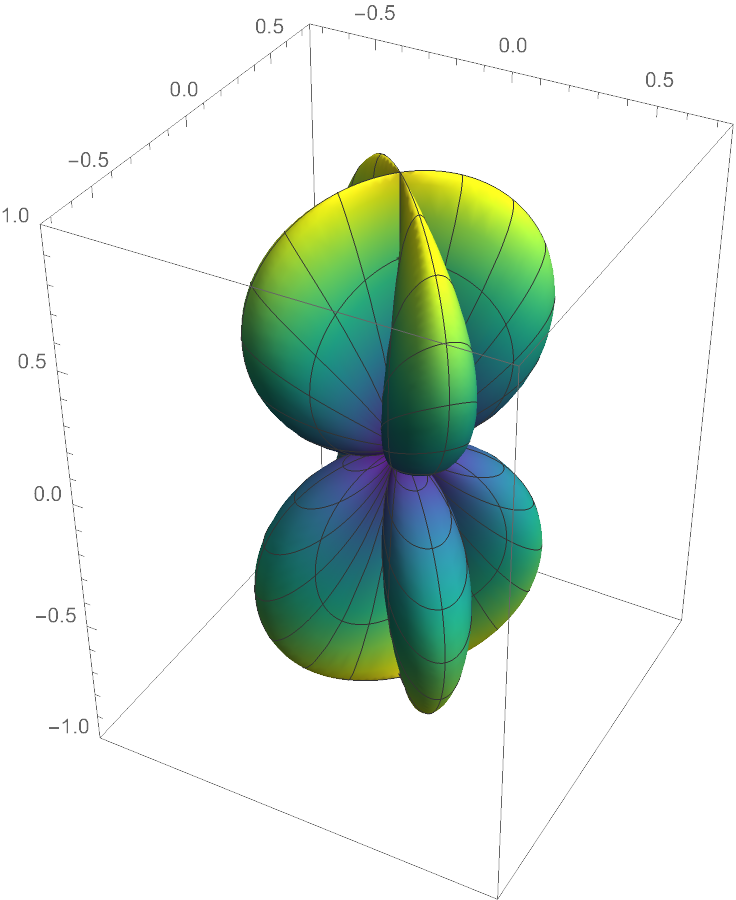}
		\subcaption{Eikonal pattern function $|F_\psi^\times|$}
		\label{fig:pattern eikonal x}
	\end{subfigure}
	\caption{Eikonal antenna pattern functions for $\Psi = 0$.
	The left panel (a) shows the antenna response pattern for a $+$ polarised \GW, the right panel (b) shows the corresponding pattern for a $\times$ polarised \GW.
	}
	\label{fig:pattern eikonal}
\end{figure}

With this parametrisation of the \GW\ polarisation, and choosing $\mI = \ee_1$ as well as $\mII = \ee_2$, the pattern functions \eqref{eq:pattern fctn phase def} evaluate to
\begin{align}
	\label{eq:pattern functions phase}
	F_\psi^+
		&= \half (1 + \cos(\Theta)^2) \cos(2 \Phi)\,,
	&
	F_\psi^\times
		&= - \cos(\Theta) \sin(2 \Phi)\,,
\end{align}
in agreement with standard results \cite[Eq.~(58)]{Yunes2013}. The absolute values of these functions are plotted in \cref{fig:pattern eikonal}, where it is seen that the \GW\ perturbation of the phase is maximal if the \GW\ propagates orthogonally to both interferometer arms.

Let us now consider the terms which are beyond the eikonal equation.

\paragraph{DC Readout Scheme}

The first corrections beyond the eikonal equation in the \DCR\ scheme are given by \cref{eq:signal DC amplitude full,eq:signal DC frequency full}. Expanding these expressions to first order in the low-frequency limit, one obtains
\begin{align}
	\label{eq:signal DC amplitude approx}
	S_{\DCR,\Aa}
	&\approx  \, \half \omegaG \ell \sin^2(\omega \Delta \ell) \left( h'(\mI, \mI) + h'(\mII, \mII) \right)\,,
	\\
	\label{eq:signal DC frequency approx}
	S_{\DCR,K}
	&\approx - \omegaG \ell \sin^2(\omega \Delta \ell) \left( h'(\mI, \mI) + h'(\mII, \mII) \right)\,,
\end{align}
where, as above, the metric perturbation is evaluated at the spatial coordinate origin.
Since the expressions coincide up to an overall factor, the corrections beyond the eikonal equation can be summarised as
\begin{align}
	\label{eq:signal DC corrections approx}
	\delta S_\DCR
	\equiv S_{\DCR,\Aa} + S_{\DCR,K}
	\approx
	- \half \omegaG \ell \sin^2(\omega \Delta \ell)
	\left( h'(\mI, \mI) + h'(\mII, \mII) \right)\,.
\end{align}
Separating the two \GW\ polarisations as in \cref{eq:GW polarisation decomposition}, this can be written as
\begin{equation}
	\delta S_\DCR
	\approx
	\omegaG \ell\, \sin^2(\omega \Delta \ell) \left[
		F_\DCR^+ h'_+ + F_\DCR^\times h'_\times
	\right]\,,
\end{equation}
where the antenna pattern functions describing the response beyond the eikonal equation in the \DCR\ scheme are
\begin{align}
	\label{eq:pattern fctn DCD def}
	F_\DCR^+
		&= - \half [A^+(m_\cI, m_\cI) + A^+(m_\cII, m_\cII)]\,,
		&
	F_\DCR^\times
		&=- \half [A^\times(m_\cI, m_\cI) + A^\times(m_\cII, m_\cII)]\,.
\end{align}
Setting $\mI = \ee_1$, $\mII = \ee_2$ and inserting the parametrisation of $A^+$ and $A^\times$ from \cref{eq:GW polarisation parametrisation}, they evaluate (for $\Psi = 0$) to
\begin{align}
	F^+_\DCR
		&= \half \sin^2(\Theta)\,,
	&
	F^\times_\DCR
		&= 0\,.
\end{align}
These pattern functions coincide (up to the overall factor $1/2$) with those of resonant bar detectors \cite[Eq.~(8.64)]{Maggiore2007}. This is because the term in parentheses in \cref{eq:signal DC corrections approx} can be written as $- h'(\ee_3, \ee_3)$ since $h$ is traceless, and resonant bar detectors are similarly only sensitive to one component of the metric perturbation. However, the expressions here contain \emph{derivatives} of $h$, while mechanical systems directly respond to the strain $h$ without differentiation.
Irrespective of this analogy, we note that the amplitude and frequency corrections are sensitive  only to $+$ polarised \GW’s (using the convention $\Psi = 0$).

\begin{figure}[ht]
	\centering
	\begin{subfigure}[c]{0.29\columnwidth}
		\includegraphics[width=\columnwidth]{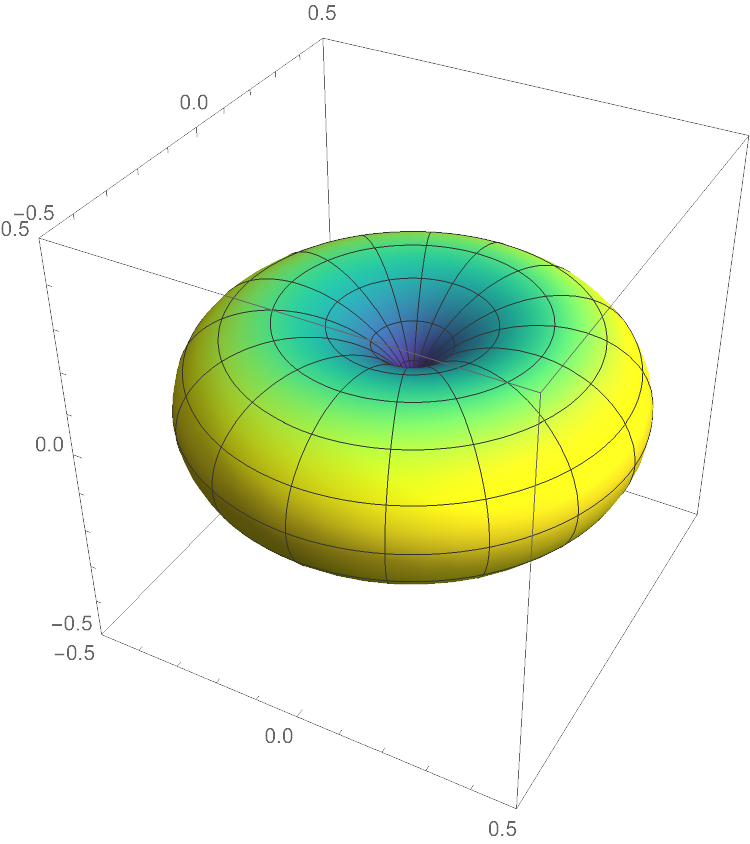}
		\subcaption{Pattern function $F_\DCR^+$}
		\label{fig:pattern DCR +}
	\end{subfigure}
	\hspace{0.05\columnwidth}
	\begin{subfigure}[c]{0.29\columnwidth}
		\includegraphics[width=\columnwidth]{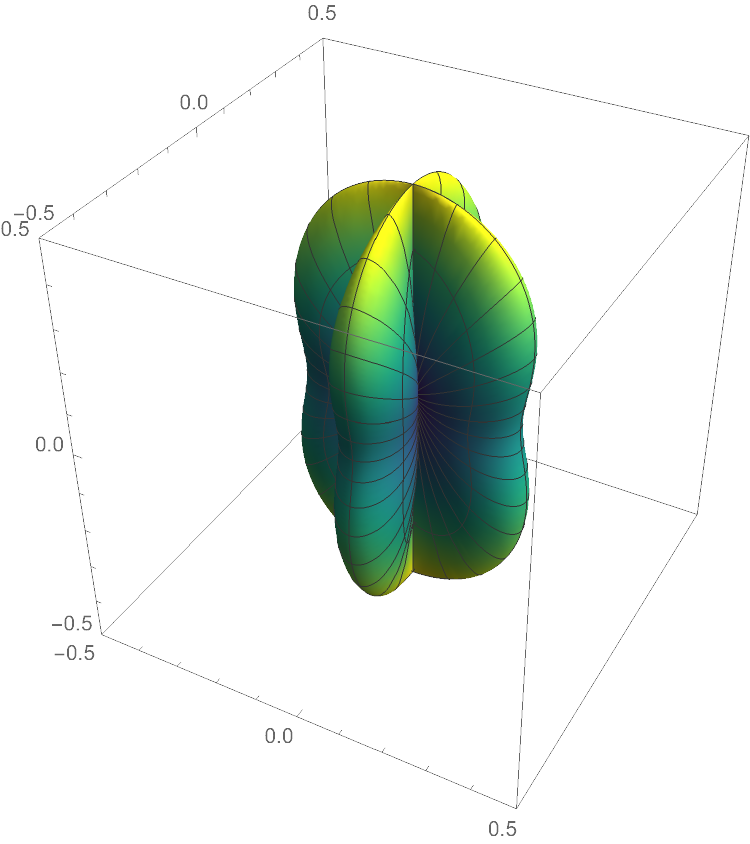}
		\subcaption{Pattern function $F_\BHR^+$}
		\label{fig:pattern BHR +}
	\end{subfigure}
	\hspace{0.05\columnwidth}
	\begin{subfigure}[c]{0.29\columnwidth}
		\includegraphics[width=\columnwidth]{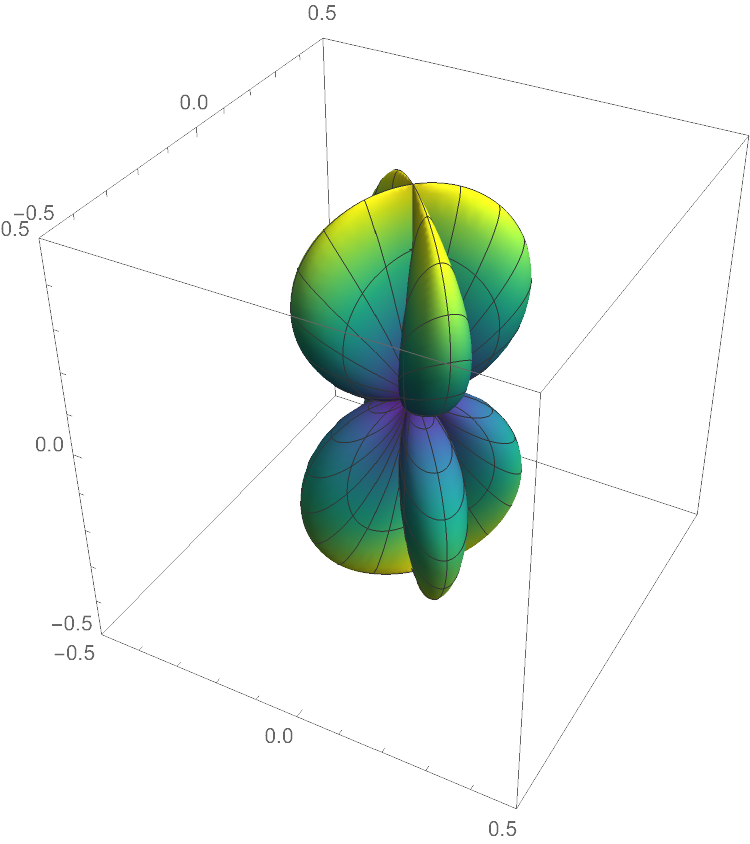}
		\subcaption{Pattern function $F_\BHR^\times$}
		\label{fig:pattern BHR x}
	\end{subfigure}
	\caption{Absolute values of the antenna pattern functions beyond the eikonal equation.
	In the DC readout scheme only $+$ polarised \GW’s contribute (a), while in the balanced homodyne readout scheme both \GW\ polarisations produce corrections to the eikonal signal (b,~c).
	}
	\label{fig:pattern functions new}
\end{figure}

The angular dependence of $F^+_\DCR$ (in absolute value) is plotted in \cref{fig:pattern DCR +}. Clearly, the effect is maximal when the \GW\ wave vector lies in the $xy$ plane (spanned by the interferometer arms), and a \GW\ propagating orthogonally to both arms produces no amplitude perturbations (contrary to the phase perturbation, which is maximal for orthogonal propagation).

\paragraph{Balanced Homodyne Readout Scheme}

The analysis of the signals in the \BHR\ scheme proceeds analogously to those in the \DCR\ scheme.
Expanding \cref{eq:signal BHR amplitude full,eq:signal BHR frequency full} to leading order in the low-frequency limit, one obtains
\begin{align}
	S_{\BHR,\Aa}
		&\approx +\quarter \omegaG \ell \cos(\varphi - 2 \omega \ell) \left( h'(\mI, \mI) - h'(\mII, \mII) \right)\,,
	\\
	S_{\BHR,K}
		&\approx -\half \omegaG \ell \cos(\varphi - 2 \omega \ell) \left( h'(\mI, \mI) - h'(\mII, \mII) \right)\,.
\end{align}
Also here, the first corrections beyond the eikonal equation can be summarised succinctly as
\begin{align}
	\delta S_\BHR
	\equiv S_{\BHR,\Aa} + S_{\BHR,K}
	\approx
		- \quarter \omegaG \ell \cos(\varphi - 2 \omega \ell) \left( h'(\mI, \mI) - h'(\mII, \mII) \right)\,,
\end{align}
and decomposing the \GW\ waveform into its two polarisation components (as above), this can be written as
\begin{equation}
	\delta S_\BHR
	\approx
	\omegaG \ell\, \cos(\varphi - 2 \omega \ell) \left[
		F_\BHR^+ h'_+ + F_\BHR^\times h'_\times
	\right]\,,
\end{equation}
where the antenna pattern functions for the response beyond the eikonal equation in the \BHR\ scheme (for $\Psi = 0$) are
\begin{align}
	F^+_\BHR
		&= -\quarter (1 + \cos(\Theta)^2) \cos(2 \Phi)\,,
	&
	F^\times_\BHR
		&= +\half \cos(\Theta) \sin(2 \Phi)\,.
\end{align}
Up to a factor $-1/2$, these functions coincide with the eikonal pattern functions given in \cref{eq:pattern functions phase}. Their absolute values are plotted in \cref{fig:pattern BHR +,fig:pattern BHR x}, for direct comparison with the pattern function of the \DCR\ scheme.

\paragraph{Angular Efficiency Factors}

To compare the sensitivity of the two readout schemes, it is customary to define the angular “sky-average” of a function $f$ as
\begin{equation}
	\overline f := \frac{1}{4 \pi}\int f(\Theta, \Phi) \sin\Theta\,\dd \Theta\, \dd \Phi\,,
\end{equation}
and to define the angular efficiency factor as
\begin{equation}
	\overline{F^2} := \overline{(F^+)^2} + \overline{(F^\times)^2}\,.
\end{equation}
For the eikonal response, and the \DCR\ and \BHR\ responses to corrections beyond the eikonal equation, these factors evaluate to
\begin{align}
	\overline{F_\psi^2}
		&= 2/5\,,
	&
	\overline{F_\DCR^2}
		&= 2/15\,,
	&
	\overline{F_\BHR^2}
		&= 1/10\,.
\end{align}
This shows that the \DCR\ scheme is more susceptible to amplitude and frequency perturbations than the \BHR\ scheme by a factor of $\sqrt{\overline{F_\DCR^2} / \overline{F_\BHR^2}} = \sqrt{4/3}$.

\paragraph{Waveforms}

\begin{figure}[ht]
	\centering
	\includegraphics[width=0.9\columnwidth]{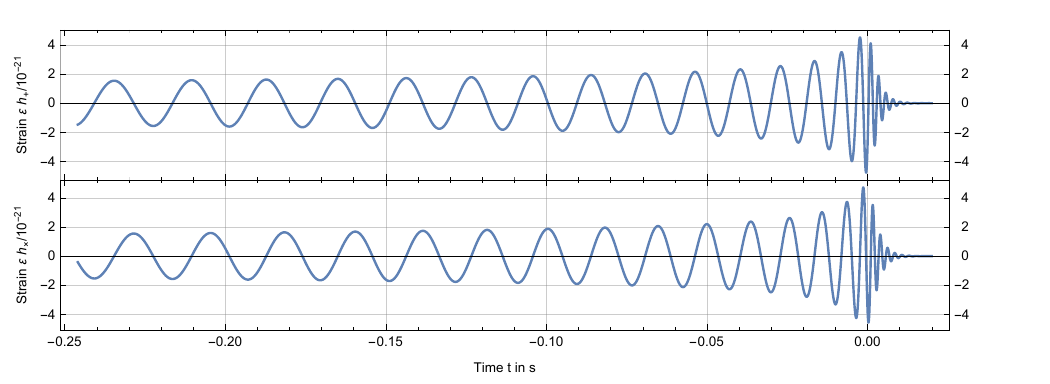}
	\caption{Exemplary waveform of a merger of two black holes of equal mass $20\,\text{M}_\odot$ at a luminosity distance of $100\,\text{Mpc}$.
	This waveform was generated from the \texttt{SEOBNRv4} model using the \texttt{PyCBC} software package.}
	\label{fig:GW waveform}
\end{figure}

To illustrate the signals produced by the first corrections beyond the eikonal equation, we plot the detector responses for the \GW\ waveform of peak amplitude $\varepsilon \approx 4.75 \times 10^{-21}$, shown in \cref{fig:GW waveform}. This waveform was generated from the \texttt{SEOBNRv4} model \cite{Bohe2017} using the \texttt{PyCBC} software package \cite{Nitz2021}.
The inspiral phase of the waveform shown here starts with a frequency of $\omegaG \approx 2\pi \times 40\,\text{Hz}$. This is to be contrasted with typical laser frequencies of $\omega \approx 2\pi \times 280\,\text{THz}$ (\textsc{ligo} currently uses a laser of $1065\,\text{nm}$ wavelength), so that the corrections beyond the eikonal equation are thus suppressed relatively to the phase signal by a factor $\omegaG/\omega \approx 1.43 \times 10^{-13}$.

\begin{figure}[ht]
	\centering
	\begin{subfigure}[c]{0.9\columnwidth}
		\includegraphics[width=\columnwidth]{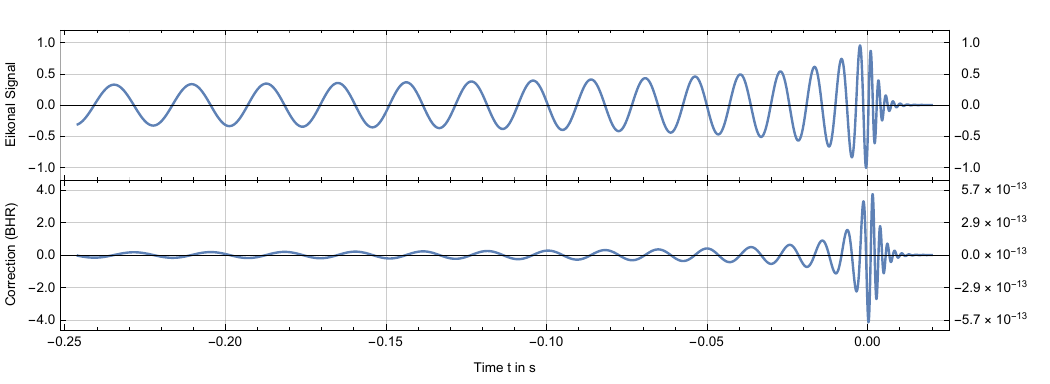}
		\subcaption{
			If the \GW\ propagates orthogonally to both interferometer arms, the phase response in both readout schemes and the beyond-eikonal response in the \BHR\ scheme are maximal, while the \DCR\ scheme is insensitive to beyond-eikonal corrections in this case (not shown).
		}
		\label{fig:GW signals normal}
	\end{subfigure}\\
	\begin{subfigure}[c]{0.9\columnwidth}
		\includegraphics[width=\columnwidth]{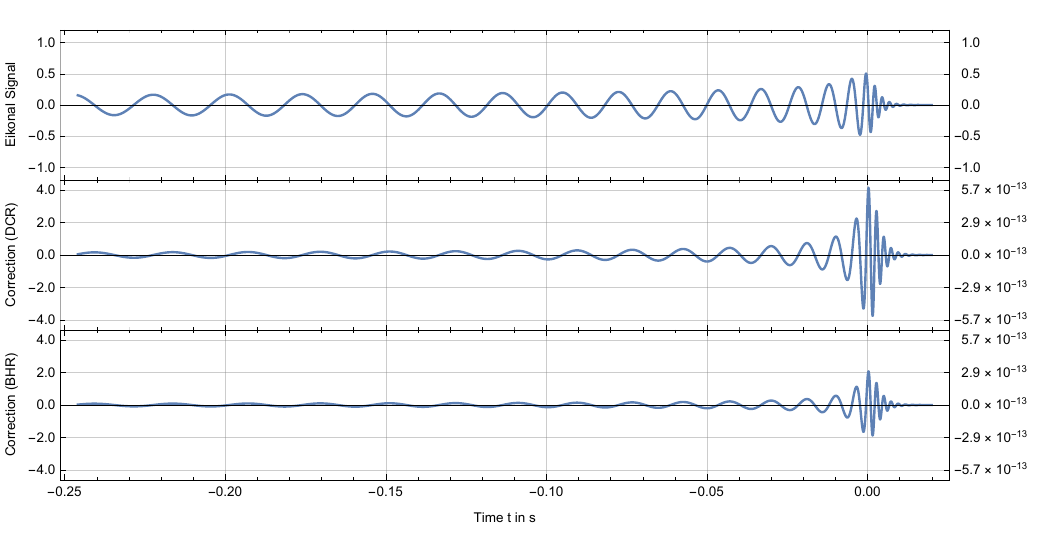}
		\subcaption{
			If the \GW\ propagates parallelly to either of the interferometer arms, both readout schemes are susceptible to corrections beyond the eikonal equation.
		}
		\label{fig:GW signals parallel}
	\end{subfigure}
	\caption{
		Comparison of the phase response and first corrections beyond the eikonal equation for the \GW\ waveform depicted in \cref{fig:GW waveform} for normal incidence (a) and parallel incidence (b).
		The left scale indicate the values of $f_\psi$, $f_\DCR$ and $f_\BHR$, while the right scale indicates the contributions to the observable signal where $f_\DCR$ and $f_\BHR$ are suppressed relatively to $f_\psi$ by the frequency ratio $\omegaG/\omega \approx 1.43 \times 10^{-13}$.
	}
	\label{fig:GW signals}
\end{figure}

Defining the dimensionless functions
\begin{align}
	\label{eq:response functions plot}
	f_\psi
		&= F_\psi^+ h_+ + F_\psi^\times h_\times\,,
	&
	f_\DCR
		&= F_\DCR^+ h'_+ + F_\DCR^\times h'_\times\,,
	&
	f_\BHR
		&= F_\BHR^+ h'_+ + F_\BHR^\times h'_\times\,,
\end{align}
the signals at leading order in the low-frequency limit can be written as
\begin{align}
	S_{\psi, \DCR}
		&\approx \omega \ell\, f_\psi \sin(2 \omega \Delta \ell)\,,
	&
	S_{\psi, \BHR}
		&\approx \omega \ell\, f_\psi \sin(\varphi - 2\omega \ell)\,,
	\\
	\delta S_{\DCR}
		&\approx \omegaG \ell\, f_\DCR \sin^2(\omega \Delta \ell)\,,
	&
	\delta S_{\BHR}
		&\approx \omegaG \ell\, f_\BHR \cos(\varphi - 2\omega \ell)\,.
\end{align}
\Cref{fig:GW signals} shows plots of the functions \eqref{eq:response functions plot} for the \GW\ waveform depicted in \cref{fig:GW waveform}. Note that these functions are normalised to a peak value of order unity (left scale), while their contributions to the detected signal differ: $f_\psi$ is multiplied by $\omega \ell$, while $f_\DCR$ and $f_\BHR$ are multiplied by $\omegaG \ell$, so that the latter two are suppressed relatively to the phase signal by $\omegaG/\omega$ (right scale).
\Cref{fig:GW signals normal} shows the signals produced for the case where the \GW\ propagates orthogonally to both interferometer arms. In this case, $S_\psi$ and $\delta S_\BHR$ are maximal, while $\delta S_\DCR$ vanishes.
For comparison, \Cref{fig:GW signals parallel} shows the signals for the case where the \GW\ propagates parallelly to one of the interferometer arms (in which  case the phase response is sensitive only to the $+$ polarisation mode). In this configuration, both the \DCR\ and the \BHR\ schemes are susceptible to beyond-eikonal corrections and, in fact, the \DCR\ scheme produces a stronger response.

However, since the corrections shown here are suppressed relatively to the phase response by the frequency ratio $\omegaG/\omega \approx 10^{-13}$, the experimental detection of such effects in the foreseeable future seems highly unlikely.

\section{Discussion}

We have determined the response of laser interferometric gravitational wave detectors at one order of accuracy beyond the eikonal equation by solving also the transport equations for the scalar amplitude and the polarisation vector. To this end, we have modelled the emission of monochromatic plane waves, described their propagation and reflection, as well as their behaviour at beam splitters.
The final result for the interferometric output intensity encompasses \GW-induced perturbations of the optical path length (via the eikonal $\psi$), modulations of the  optical frequency (as given by the time-component of the local wave vectors $K$), and scintillations (described by the amplitudes $\Aa$), but was found to be independent of the perturbation of the polarisation vector $E$ arising from a non-trivial holonomy integral along the light rays.

The analysis presented here goes beyond previous results which have either allowed for arbitrary \GW\ incidence angles but were restricted to the level of the eikonal equation \cite{Montanari1998}, or have described amplitude and polarisation effects but were restricted to special alignments \cites{Cooperstock1993,Calura1999,Lobato2021}. Moreover, the analysis here does not rely on unidimensional approximations such as in Refs.~\cites{Lobo1992,Tarabrin2007}, and the result is well-behaved also for small angles between the \GW\ and \EM\ wave vectors, where calculations based on the vector potential often find difficulties \cite{Park2021}.

Finally, we comment on the model assumptions made here.
The geometry of the interferometer was assumed to be well-described in terms of the background geometry (when using transverse-traceless coordinates), which relies on the assumption that all material points of the interferometer follow geodesic motion. This will not hold exactly true in experiment, and deviations from this model would require detailed descriptions of the elastic properties of the respective materials. However, since the mirrors and beam splitters are much smaller than the interferometer arm lengths, we expect such corrections to be suppressed by the ratio of the object size compared to the arm length and thus negligible for most applications.
Further, we have made specific assumptions of the light sent into the interferometer. While the assumptions on the eikonal and amplitude are simple to interpret (emission occurs with definite and fixed frequency and the emitted radiation has constant energy density), the emitted polarisation is more difficult to model, leading to the unspecified term $E'$ in \cref{eq:emitted polarisation}. However, the final result is independent of $E'$ due to it being orthogonal to the unperturbed polarisation $E^\o0$. Hence, we expect most other models to differ mainly in the choice of the emission surfaces on which the boundary conditions are prescribed, rather than in the data prescribed thereon. However, provided the surfaces differ from the ones used here only at first order in $\varepsilon$, such arising corrections can be described entirely using the unperturbed transport equations since corrections to the latter would produce terms quadratic in $\varepsilon$.

We expect that the calculations presented here can be generalised without much complication to Fabry-Pérot cavities, while optical cavities whose mirrors are not plane-parallel (but rather concentric or confocal, for example) could be described using similar techniques, but would require different boundary conditions adapted to the problem.

\section*{Acknowledgements}
I thank Piotr Chruściel, Stefan Palenta and Christopher Hilweg for helpful discussions and gratefully acknowledge funding via a fellowship of the Vienna Doctoral School in Physics (VDSP) and the Austrian Science Fund (FWF) through GRIPS (Grant No. P30817-N36), as well as support by TURIS.

\clearpage
\appendix
\section{Derivation of the Polarisation Transport Equations}
\label{appendix:geometrical optics}

The equations of geometrical optics can be motivated by writing the electromagnetic field strength tensor as
\begin{equation}
    \label{eq:go ansatz}
	\t F{_\mu_\nu} = \t {\mathfrak f}{_\mu_\nu} e^{i \omega \psi}\,,
\end{equation}
and inserting for $\t {\mathfrak f}{_\mu_\nu}$ a \emph{formal} expansion in inverse powers of $(i \omega)$, see e.g.\ Refs.~\cites[Sect.~4.1]{Kravtsov1990}[Sect.~2.4]{Perlick2000}{Ehlers1967}.
Using \cref{eq:go ansatz}, Maxwell’s equation in vacuum take the form
\begin{align}
	\label{eq:go first order}
	i \omega \dd \psi \wedge \mathfrak f + \dd \mathfrak f &= 0\,,
	&
	i \omega \nabla \psi \cdot \mathfrak f + \nabla \cdot \mathfrak f &= 0\,,
\end{align}
where a dot indicates contraction of adjacent indices.
Together, these equations imply
\begin{equation}
	i \omega |\nabla \psi|^2 \mathfrak f
	+ \nabla \psi \cdot \mathfrak \dd \mathfrak f
	+ \dd \psi \wedge (\nabla \cdot \mathfrak f)
	= 0\,.
\end{equation}
Assuming $\psi$ to satisfy the eikonal equation
\begin{equation}
	|\nabla \psi|^2
	\equiv \t g{^\mu^\nu} (\t\nabla{_\mu} \psi) (\t\nabla{_\nu} \psi)
	= 0\,,
\end{equation}
one is left with the intermediate result
\begin{equation}
	\label{eq:go consistency}
	\nabla \psi \cdot \mathfrak \dd \mathfrak f
	+ \dd \psi \wedge (\nabla \cdot \mathfrak f)
	= 0\,.
\end{equation}
Further, taking the divergence of the first equation in \eqref{eq:go first order} and adding the exterior derivative of the second yields
\begin{equation}
	\label{eq:go intermediate}
	\dd (\nabla \psi \cdot \ff) + \nabla \cdot(\dd \psi \wedge \ff)
	= (i \omega)^{-1} \wop_\text{HdR} \ff \,,
\end{equation}
where $\wop_\text{HdR} \ff = - \dd (\nabla \cdot \ff) - \nabla \cdot (\dd \ff)$ is the Hodge-de~Rham wave operator, which is related to the connection-induced wave operator by the Weitzenböck identity
\begin{equation}
	\wop_\text{HdR} \t\ff{_\mu_\nu}
		= - \t\nabla{^\alpha} \t\nabla{_\alpha} \t\ff{_\mu_\nu}
		+ \t g{^\alpha^\beta}\left(
			\t R{^\sigma_\beta_\mu_\alpha} \t \ff{_\sigma_\nu}
			+\t R{^\sigma_\nu_\mu_\alpha} \t \ff{_\beta_\sigma}
			-\t R{^\sigma_\beta_\nu_\alpha} \t \ff{_\sigma_\mu}
			-\t R{^\sigma_\mu_\nu_\alpha} \t \ff{_\beta_\sigma}
		\right)\,,
\end{equation}
where $\t R{^\alpha_\beta_\mu_\nu}$ denotes the Riemann curvature tensor.
A direct calculation shows
\begin{equation}
	\dd (\nabla \psi \cdot \ff) + \nabla \cdot(\dd \psi \wedge \ff)
	= 2 \nabla_{\nabla \psi} \ff + (\wop \psi) \ff - \left(
		\dd \psi \wedge (\nabla \cdot \ff)
		+ (\nabla \psi) \cdot \dd \ff
	\right)\,,
\end{equation}
where the last term vanishes by virtue of \cref{eq:go consistency}.
\Cref{eq:go intermediate} thus yields
\begin{equation}
	\label{eq:go second order}
	2 \nabla_{\nabla \psi} \ff + (\wop \psi) \ff
	= (i \omega)^{-1} \wop_\text{HdR} \ff\,,
\end{equation}
from which the transport equations of geometrical optics can be obtained by inserting the formal expansion
\begin{equation}
	\label{eq:go amplitude expansion}
	\ff
		= \sum_{m = 0}^\infty (i \omega)^{-m}\, \ff_\o{m}\,,
\end{equation}
which yields
\begin{align}
	2 \nabla_{\nabla \psi} \ff_\o{0} + (\wop \psi) \ff_\o{0} &= 0\,,
	&
	2 \nabla_{\nabla \psi} \ff_\o{m+1} + (\wop \psi) \ff_\o{m+1} &= \wop_\text{HdR} \ff_\o{m}\,.
\end{align}
Substituting \eqref{eq:go amplitude expansion} in the first-order equations \eqref{eq:go first order} then yields the algebraic recursion relations
\begin{align}
	\nabla \psi \cdot \ff_\o{0} &= 0\,,
	& \nabla \psi \cdot \ff_\o{m+1} &= \nabla \cdot \ff_\o{m}\,,
	\\
	\dd \psi \wedge \ff_\o{0} &= 0\,,
	& \dd \psi \wedge \ff_\o{m+1} &= \dd \ff_\o{m}\,,
\end{align}
which can be thought of as constraint equations limiting the number of degrees of freedom to two.
The geometrical optic scheme described here is in accord with the one given in Ref.~\cite{Preti2010}.

We emphasise that the series \eqref{eq:amplitude transport} is first and foremost a \emph{formal} series, which is not necessarily convergent, but rather asymptotic in many typical applications, see e.g.\ Ref.~\cite[Sect.~2.7]{Perlick2000} for a discussion of the asymptotic nature of such series for initial data prescribed on spacelike hypersurfaces.
In the concrete setup here, our analysis of the full Einstein-Maxwell equations (in preparation) will assess the quality of the leading order geometrical optics approximation and estimate the errors made in using this approximation.

\section{Interpretation of the Polarisation Vector}
\label{s:polarisation interpretation}

In this section, we show how the one-form $E$ in the field strength tensor
\begin{equation}
	\label{eq:F for interpretation}
	F = \Aa e^{i \omega \psi} (\dd \psi) \wedge E
\end{equation}
is related to the polarisation of the electromagnetic wave.
Here, $E$ is actually only defined up to addition of a multiple of $\dd \psi$.

Consider an observer $\Oo$ with four-velocity $u$ (i.e.\ a unit timelike vector).
Since the eikonal $\psi$ describes the rapidly oscillating phase of the electromagnetic wave, the measured frequency is
\begin{equation}
	\label{eq:observed frequency}
	\omega_\Oo = - \omega \,\t u{^\mu} \t \nabla{_\mu} \psi\,,
\end{equation}
so that $-\t u{^\mu} \t \nabla{_\mu} \psi = \omega_\Oo/\omega$.
Regarding the electric field $\Ee$ and the magnetic field $\Bb$ as one- and two-forms, respectively, the decomposition of the field strength can be written as
\begin{equation}
	\label{eq:F decomposition E B}
	\t F{_\mu_\nu}
		= 2 \t u{_[_\mu} \t \Ee{_\nu_]}
		+ \t \Bb{_\mu_\nu}\,,
\end{equation}
where $\Ee$ and $\Bb$ are such that their contractions with $u$ vanish (i.e.\ they are spatial for $\Oo$). In particular, they are determined by
\begin{align}
	\label{eq:observed fields E B}
	\t\Ee{_\mu}
		&= - \t u{^\nu} \t F{_\nu_\mu}\,,
	&
	\t\Bb{_\mu_\nu}
		&= \t F{_\mu_\nu} - 2 \t u{_[_\mu} \t \Ee{_\nu_]}\,.
\end{align}
We now apply this decomposition to the field given in \cref{eq:F for interpretation}. Using the freedom to add any multiple of $\dd \psi$ to $E$, we write
\begin{align}
	F
		&= \Aa e^{i \omega \psi} (\dd \psi) \wedge \bar E\,,
	\\
\shortintertext{where}
	\t {{\bar E}}{_\mu}
		&= \t E{_\mu} -  \frac{\t u{^\alpha} \t E{_\alpha}}{\t u{^\beta} \t\nabla{_\beta} \psi} \t \nabla{_\mu} \psi\,,
\end{align}
which is chosen such that $\t u{^\mu} \t {{\bar E}}{_\mu} = 0$. The above formula for $\Ee$ then yields
\begin{equation}
	\Ee = (\Aa  \omega_\Oo/\omega)\, \bar E e^{i \psi} \,,
\end{equation}
so that $\bar E$ determines the direction of the electric field and $\Aa \omega_\Oo / \omega$ its amplitude (note that $\bar E$ has the same norm as $E$). To obtain an expression for the magnetic field $\Bb$, we first decompose the eikonal gradient as
\begin{equation}
	\t \nabla{_\mu} \psi = \tfrac{\omega_\Oo}{\omega}(\t u{_\mu} + \t n{_\mu} )\,,
\end{equation}
where $\t n{_\mu}$ is a unit spacelike vector orthogonal to $u$.
Using the second equation in \eqref{eq:observed fields E B}, one finds
\begin{equation}
	\label{eq:nEB frame}
	\Bb = n \wedge \Ee\,,
\end{equation}
which is the standard formula stating that $\Ee, \Bb$ and $n$ form a right-handed orthogonal system, and that the electric and magnetic fields have the same norm.

\section{Constraint Equations for Boundary Data}
\label{appendix:constraints}

As is well-known from the $3+1$ decomposition of Maxwell’s equations, one is not free to prescribe the electromagnetic field on spacelike hypersurfaces arbitrarily.
Instead, the initially prescribed electric and magnetic fields $\Ee$ and $\Bb$ (in Heaviside–Lorentz units) must satisfy $\Dd \cdot \Ee = \rho$ and $\Dd \cdot \Bb = 0$, where $\rho$ is the charge density on the initial hypersurface, and $\Dd$ is the spatial covariant derivative.

Similarly, we show here that boundary values prescribed on a \emph{timelike} hypersurface $\Sigma$ must also satisfy certain constraint equations, which we compute explicitly for the case of normal emission, as considered in the main body of the text.

To this end, consider a hypersurface-orthogonal unit timelike vector field $u$. With respect to this field, one may decompose the field strength, $F$, as well as its Hodge dual, $\ast F$, in terms electric and magnetic fields as
\begin{align}
	\t F{^\mu^\nu}
		&= + \t u{^\mu} \t\Ee{^\nu}
		- \t u{^\nu} \t\Ee{^\mu}
		+ \t \epsilon{^\mu^\nu^\rho} \t \Bb{_\rho}\,,
	&
	\t{\ast F}{^\mu^\nu}
		&= - \t u{^\mu} \t \Bb{^\nu}
		+ \t u{^\nu} \t \Bb{^\mu}
		+ \t \epsilon{^\mu^\nu^\rho} \t \Ee{_\rho}\,,
\end{align}
where $\Ee$ and $\Bb$ are both spatial in the sense that their contraction with $u$ vanishes, cf.~\cref{eq:F decomposition E B}.
Maxwell’s equations in vacuum then take the form
\begin{align}
    \label{eq:Maxwell 3+1 E}
	\LieD_u \t \Ee{^\mu}
	+ (\t \nabla{_\nu} \t u{^\nu}) \t\Ee{^\mu}
	- (\t \nabla{_\nu} \t \Ee{^\nu}) \t u{^\mu}
	- \t \epsilon{^\alpha^\mu^\nu^\rho} \t \nabla{_\nu} \t u{_\alpha} \t \Bb{_\rho}
	- \t \epsilon{^\mu^\nu^\rho} \t\nabla{_\nu} \t \Bb{_\rho}
    &= 0\,,
	\\
    \label{eq:Maxwell 3+1 B}
	\LieD_u \t \Bb{^\mu}
	+ (\t \nabla{_\nu} \t u{^\nu}) \t\Bb{^\mu}
	- (\t \nabla{_\nu} \t \Bb{^\nu}) \t u{^\mu}
	+ \t \epsilon{^\alpha^\mu^\nu^\rho} \t \nabla{_\nu} \t u{_\alpha} \t \Ee{_\rho}
	+ \t \epsilon{^\mu^\nu^\rho} \t\nabla{_\nu} \t \Ee{_\rho}
    &= 0\,,
\end{align}
where $\LieD_u$ denotes the Lie derivative along the vector field $u$.

Consider now a timelike hypersurface $\Sigma$ with unit normal $n$, such that $u$ and $n$ are orthogonal. This entails that $n$ is spatial with respect to $u$, and also that $u$ is tangent to $\Sigma$. Contracting \cref{eq:Maxwell 3+1 E,eq:Maxwell 3+1 B} with the surface normal $n$ and imposing normal emission in the sense
\begin{align}
	\label{eq:normal components EB}
	\t \Ee{_\mu} \t n{^\mu} &= 0\,,
	&
	\t \Bb{_\mu} \t n{^\mu} &= 0\,,
\end{align}
one obtains
\begin{align}
	\t{\LieD_u \Ee}{^\mu} \t n{_\mu}
	- \t \epsilon{^\mu^\nu^\rho} \t n{_\mu} \t\nabla{_\nu} \t \Bb{_\rho}
	- \t \epsilon{^\mu^\nu^\rho} \t n{_\mu} \t a{_\nu} \t \Bb{_\rho}
	&= 0\,,
	\\
	\t{\LieD_u \Bb}{^\mu} \t n{_\mu}
	- \t \epsilon{^\mu^\nu^\rho} \t n{_\mu} \t\nabla{_\nu} \t \Ee{_\rho}
	- \t \epsilon{^\mu^\nu^\rho} \t n{_\mu} \t a{_\nu} \t \Ee{_\rho}
	&= 0\,,
\end{align}
where $a = \nabla_u u $ is the acceleration vector. These equations are constraints on boundary data, as they only involve tangential derivatives, but no derivatives normal to $\Sigma$.

It turns out that the first terms involving $\LieD_u$ of $\Ee$ and $\Bb$ give no contribution.
Indeed, writing the Lie derivative in terms of the Lie bracket $\LieD_u \Ee = [u, E]$, and noting that $u$ and $E$ being tangent to $\Sigma$ implies that their Lie bracket is also tangent to $\Sigma$, one finds $\t{\LieD_u \Ee}{^\mu} \t n{_\mu} = \t{\LieD_u \Bb}{^\mu} \t n{_\mu} = 0$.
Hence, the constraint equations reduce to
\begin{align}
    \label{eq:constraint normal emission}
	\t \epsilon{^\mu^\nu^\rho} \t n{_\mu} (\t\nabla{_\nu} \t \Bb{_\rho} + \t a{_\nu} \t \Bb{_\rho})
	&= 0\,,
	&
	\t \epsilon{^\mu^\nu^\rho} \t n{_\mu} (\t\nabla{_\nu} \t \Ee{_\rho} + \t a{_\nu} \t \Ee{_\rho})
	&= 0\,,
\end{align}
which are the same for both $\Ee$ and $\Bb$.
Assuming further (as is the case for plane waves) that $n, \Ee, \Bb$ form a right-handed orthogonal frame with $|\Ee| = |\Bb|$, such that \cref{eq:nEB frame} applies, one may eliminate one of the fields from the constraints.
In this case, one has
\begin{equation}
	\t \epsilon{^\mu^\nu^\rho} \t n{_\mu} \t\nabla{_\nu} \t \Bb{_\rho}
	= (\t*\delta{^\mu_\nu} + \t u{^\mu} \t u{_\nu} - \t n{^\mu} \t n{_\nu}) \t\nabla{_\mu} \t \Ee{^\nu}\,,
\end{equation}
so that the first equation of \eqref{eq:constraint normal emission} takes the form
\begin{equation}
	(\t*\delta{^\mu_\nu} + \t u{^\mu} \t u{_\nu} - \t n{^\mu} \t n{_\nu}) \t\nabla{_\mu} \t \Ee{^\nu}
	+ \t a{^\mu} \t \Ee{_\mu} = 0\,.
\end{equation}
Writing
\begin{align}
    \tilde \delta E &= (\t*\delta{^\mu_\nu} + \t u{^\mu} \t u{_\nu} - \t n{^\mu} \t n{_\nu}) \t\nabla{_\mu} \t \Ee{^\nu}\,,
\end{align}
where $\tilde \delta$ is the divergence operator of the intersection of $\Sigma$ with a time-slice, and similarly denoting the exterior derivative intrinsic to such a surface by $\tilde \dd$, one obtains the constraint equations in the form
\begin{align}
	\tilde \delta \Ee + a_\parallel \cdot \Ee &= 0\,,
	&
	\tilde \dd \Ee + a_\parallel \wedge \Ee &= 0\,,
\end{align}
where $a_\parallel = a - n (n\cdot a)$ is the tangential part of the acceleration, and where we have identified the vector $a$ with its metric-equivalent one-form.
In particular, for geodesic $u$, this simplifies to
\begin{align}
	\tilde \delta \Ee &= 0\,,
	&
	\tilde \dd \Ee & = 0\,,
\end{align}
which entails that $\Ee$ satisfies $\Delta_\text{HdR} \Ee = 0$, where $\Delta_\text{HdR}$ is the Hodge-de~Rham Laplacian. Moreover, on a contractible region the second equation implies that $\Ee$ is exact so that the problem reduces to the scalar Laplace equation
\begin{align}
	\Ee &= \dd \chi\,,
	&
	&\text{where}
	&
	\Delta \chi &= 0\,.
\end{align}
Here, $\Delta$ denotes the spatial scalar Laplacian on $\Sigma$. Note that $\chi$ is not required to remain bounded: instead, since the physically relevant field is $\Ee = \dd \chi$, we require $\chi$ to have bounded first derivatives.

To construct perturbative solutions, it is convenient to write the Laplace equation using the “densitised contravariant metric” in the form
\begin{align}
	\t\p{_A} (\t{\tilde{\mathfrak g}}{^A^B} \t\p{_B} \chi) = 0\,,
\end{align}
where $\t{\tilde{\mathfrak g}}{^A^B} = \sqrt{\det \tilde g}\,\t{\tilde g}{^A^B}$, with $\tilde g$ being the metric induced by $g$ on $\Sigma$ (at any instant of time). Inserting the perturbative expansion
\begin{equation}
	\chi = \t E{^{\o0}_A} \t x{^A} + \varepsilon \chi^\o1\,,
\end{equation}
one obtains
\begin{equation}
	\Delta^\o0 \chi^\o1
	+ \t\p{_A} \t{\tilde{\mathfrak g}}{^{\o1}^A^B} \t E{^{\o0}_B} = 0\,,
\end{equation}
where $\Delta^\o0 = \t \delta{^A^B} \t\p{_A} \t\p{_B}$ is the unperturbed Laplacian on $\Sigma$.
Since $\tilde{\mathfrak g}$ is a function of $\kappa.x$ alone, one finds a particular solution to be given by
\begin{equation}
	\chi^\o1_\text{part.}
		= - \frac{\t\kappa{_B} \t E{^{\o0}_C}}{{\tilde{\mathfrak g}}^\o0(\kappa, \kappa)}
		\int_0^{\kappa.x} \t{\tilde{\mathfrak g}}{^{\o1}^B^C}(u) \,\dd u\,,
\end{equation}
and requiring the overall function $\chi$ to have bounded first derivatives, one is only free to add homogeneous solutions of the form $\t\chi{_A} \t x{^A}$, where the $\t\chi{_A}$’s are constant.
Taking the derivative of
\begin{equation}
	\chi^\o1
		= - \frac{\t\kappa{_B} \t E{^{\o0}_C}}{{\tilde{\mathfrak g}}^\o0(\kappa, \kappa)}
		\int_0^{\kappa.x} \t{\tilde{\mathfrak g}}{^{\o1}^B^C}(u) \,\dd u
		+ \t\chi{_A} \t x{^A}
		\,,
\end{equation}
and choosing the constants $\t\chi{_A}$ such that $E^\o1$ assumes a prescribed value at the spatial origin, one obtains
\begin{equation}
	\t E{^{\o1}_A}
		= \t E{^{\o1}_A}(t, \t x{^i} = 0)
		+ \frac{\t\kappa{_A} \t\kappa{_B} \t E{^{\o0}_C}}{{\tilde{\mathfrak g}}^\o0(\kappa, \kappa)} \left[
			\t{\tilde{\mathfrak g}}{^{\o1}^B^C}(\omega t)
			- \t{\tilde{\mathfrak g}}{^{\o1}^B^C}(\kappa.x)
		\right]\,.
\end{equation}
\section{Boundary Conditions for Perfect Reflection}
\label{appendix:perfect reflection}

In this section, we give a generally covariant description of perfect reflection at the level of geometrical optics.
For this, we consider a field of the form
\begin{align}
	F
		&= F_1 + F_2\,,
	&
	F_{1,2}
		&= (\dd \psi_{1,2}) \wedge E_{1,2} \, e^{i \omega \psi_{1,2}}\,,
\end{align}
where $F_1$ describes a wave impinging on a reflecting surface $\Sigma$, and $F_2$ is the reflected wave.
We first consider how $\dd \psi_2$ relates to $\dd \psi_1$ (on $\Sigma$), and then we determine $E_2$ from $E_1$.

For the eikonal functions, we assume the standard matching condition that $\psi_1$ and $\psi_2$ coincide on $\Sigma$, cf.\ e.g.\ Ref.~\cite[Sect.~2.5]{Kravtsov1990}.
This entails that $\dd \psi_2 = \dd \psi_1 + \alpha \nu$ (on $\Sigma$), where $\nu$ is the unit conormal to $\Sigma$, and $\alpha$ is some function on $\Sigma$ which can be determined as follows. Since we are concerned with waves propagating in different directions, we must have $\alpha \neq 0$. Further, taking the norm on both sides and using the fact that both functions $\psi_{1,2}$ satisfy the eikonal equation, one obtains $\alpha = - 2 \t \nu{^\sigma} \t \nabla{_\sigma} \psi_1$. Using the definition \eqref{eq:reflection tensor}, this can be written as
\begin{equation}
	\label{eq:reflected wave vector}
	\t \nabla{_\rho} \psi_2 = \t R{_\rho^\sigma} \t \nabla{_\sigma} \psi_1\,,
\end{equation}
i.e.\ the wave vector is simply reflected along the normal to $\Sigma$.

We now turn to the relationship between $E_1$ and $E_2$.
For this, consider an arbitrary point on $\Sigma$ (which we shall keep fixed from now on) and let $u$ denote any (future-pointing) unit timelike vector at this point, which is orthogonal to $\nu$ (the final result will be independent of this choice). By the argument of \cref{s:polarisation interpretation}, we may assume without loss of generality that the covectors $E_{1,2}$ are orthogonal to $u$.
Using local inertial coordinates adapted to $u$ (at the considered point), we may write
\begin{equation}
	\dd \psi_{1,2} = \frac{\omega_\Oo}{\omega} (m_{1,2} - \dd t)\,,
\end{equation}
where $\omega_\Oo$ is the frequency measured by observers with four-velocity $u$, cf.\ \eqref{eq:observed frequency}, and where $m_{1,2}$ are spatial unit covectors which are related by
$\t {m}{_2_\rho} = \t R{_\rho^\sigma}\t {m}{_1_\sigma}$.
On $\Sigma$, the field strength then evaluates to
\begin{equation}
	F
		= \left[
			- \dd t \wedge (E_1 + E_2)
			+ m_1 \wedge E_1 + m_2 \wedge E_2
		 \right] \tfrac{\omega_\Oo}{\omega} e^{i \omega \psi}\,,
\end{equation}
and comparison with \cref{eq:F decomposition E B} yields
\begin{align}
	\Ee
		&= (E_1 + E_2) \tfrac{\omega_\Oo}{\omega} e^{i \omega \psi}\,,
	&
	\Bb
		&= (m_1 \wedge E_1 + m_2 \wedge E_2) \tfrac{\omega_\Oo}{\omega} e^{i \omega \psi}\,.
\end{align}
Now, since we are working in a local inertial frame, we may use the standard equations
\begin{align}
	\vec \nu \times \vec\Ee &= 0\,,
	&
	\vec \nu \cdot \vec\Bb &= 0\,,
\end{align}
which are equivalent to the following equations, in exterior algebra notation:
\begin{align}
	\label{eq:reflection conditions intermediate}
	\nu \wedge \Ee &= 0\,,
	&
	\nu \wedge \Bb &= 0\,.
\end{align}
The first equation here entails that $E_2 = - E_1 + \beta \nu$ for some constant $\beta$, and the condition of perfect reflection requires that $E_1$ and $E_2$ have the same norm, yielding
\begin{equation}
	\t E{_2_\rho} = - \t R{_\rho^\sigma} \t E{_1_\sigma}\,.
\end{equation}
The second matching condition $\nu \wedge \Bb =  0$ is then automatically satisfied. Indeed, since ${\nu \wedge m_2 = \nu \wedge m_1}$, one has
\begin{equation}
	\nu \wedge \Bb
		= \nu \wedge m_1 \wedge \Ee\,,
\end{equation}
which vanishes because $\nu \wedge \Ee = 0$.

So far, we have shown that in local inertial coordinates (using a decomposition of $F$ adapted to the four-velocity $u$) the conditions for perfect reflection are
\begin{align}
	\t \nabla{_\rho} \psi_2 &= \t R{_\rho^\sigma} \t \nabla{_\sigma} \psi_1 \,,
	&
	\t E{_2_\rho} &= - \t R{_\rho^\sigma} \t E{_1_\sigma} \,,
\end{align}
which, together, imply
\begin{equation}
	\t F{_2_\alpha_\beta}
		= - \t R{_\alpha^\rho} \t R{_\beta^\sigma} \t F{_1_\rho_\sigma}\,.
\end{equation}
But since this last equation (which is independent of the decomposition of $F$ into $\Ee$ and $\Bb$) is tensorial, it holds in all coordinate systems, and since the considered point on $\Sigma$ was arbitrary, it holds everywhere on this surface.

\section{Output Power of an Interferometer}
\label{appendix:output power}

In \cref{eq:detector field}, the field at the output port of the interferometer was found to be of the form
\begin{equation}
	\Ff
		= \Aa_\cI\, k_\cI \wedge E_\cI\, \cos(\psi_\cI)
		+ \Aa_\cII\, k_\cII \wedge E_\cII\, \cos(\psi_\cII)\,,
\end{equation}
where $k_\cI$ and $k_\cII$ are null vectors, and $E_\cI$ and $E_\cII$ are spacelike unit vectors (with respect to the full space-time metric $g$), which satisfy $g(k_\cI, E_\cI) = 0$ and $g(k_\cII, E_\cII) = 0$.
In this section, we compute the time-averaged energy density for such a field at first order in the \GW\ amplitude $\varepsilon$.

The intensity measured by a photodetector is given by the temporal component of the energy-momentum tensor
\begin{equation}
	\t T{_\mu_\nu}
		= \t \Ff{_\mu_\alpha} \t \Ff{_\mu_\beta} \t g{^\alpha^\beta}
		- \quarter \t g{_\mu_\nu}\, \t \Ff{^\alpha^\beta} \t \Ff{_\alpha_\beta}\,.
\end{equation}
Considering the first term only for the moment, we have
\begin{equation}
\begin{split}
	\t \Ff{_0_\alpha} \t \Ff{_0_\beta} \t g{^\alpha^\beta}
		&= (\Aa_\cI\, \t k{_\cI_0})^2 |E_\cI|^2 \cos(\psi_\cI)^2
		+ (\Aa_\cII\, \t k{_\cII_0})^2 |E_\cII|^2 \cos(\psi_\cII)^2
		\\&
		+ 2 \Aa_\cI\, \Aa_\cII\,  \left(
			\t k{_\cI_0} \t k{_\cII_0} g(E_\cI, E_\cII)
			+ \t E{_\cI_0} \t E{_\cII_0} g(k_\cI, k_\cII)
			\right) \cos(\psi_\cI) \cos(\psi_\cII)
		\\&
		- 2 \Aa_\cI\, \Aa_\cII \left(
			\t k{_\cI_0} \t E{_\cII_0} g(E_\cI, k_\cII)
			+ \t k{_\cII_0} \t E{_\cI_0} g(E_\cII, k_\cI)	
		\right) \cos(\psi_\cI) \cos(\psi_\cII)
	\,,
\end{split}
\end{equation}
which can be simplified considerably.
Since $E_\cI$ and $E_\cII$ have unit norm, the first line simplifies to $(\Aa_\cI\, \t k{_\cI_0})^2 \cos(\psi_\cI)^2 + (\Aa_\cII\, \t k{_\cII_0})^2 \cos(\psi_\cII)^2$.
For the remaining expression, we make use of the fact that in the unperturbed problem the $k$’s coincide, and so do the $E$’s.

The normalisation conditions imply
\begin{align}
	E^\o0 . E_\cI^\o1 &= \half h(E^\o0, E^\o0)\,,
	&
	k^\o0 . k_\cI^\o1 &= \half h(k^\o0, E^\o0)\,,
\end{align}
and similarly for $E_\cII^\o1$ and $k_\cII^\o1$. This implies
\begin{align}
	\label{eq:energy inner producs}
	g(E_\cI, E_\cII) &= 1 + O(\varepsilon^2)\,,
	&
	g(k_\cI, k_\cII) &= 0 + O(\varepsilon^2)\,,
\end{align}
so that the second line above reduce to $2 (\Aa_\cI \t k{_\cI_0}) (\Aa_\cII \t k{_\cII_0}) \cos(\psi_\cI) \cos(\psi_\cII)$ at the considered order.

Moreover, the inner products $g(E_\cI, k_\cII)$ and $g(E_\cII, k_\cI)$ in the third line are of order $\varepsilon$ since the unperturbed wave vectors are orthogonal to the unperturbed polarisation vectors. But since the unperturbed polarisation vectors are purely spatial, the multiplying factors $\t E{_\cI_0}$ and $\t E{_\cII_0}$ are of order $\varepsilon$ as well, so that both summands in the last line are of second order and thus negligible.

Hence, to first order in $\varepsilon$ one has
\begin{equation}
\begin{split}
	\t \Ff{_0_\alpha} \t \Ff{_0_\beta} \t g{^\alpha^\beta}
		&= (\Aa_\cI\, \t k{_\cI_0})^2 \cos(\psi_\cI)^2
		+ (\Aa_\cII\, \t k{_\cII_0})^2 \cos(\psi_\cII)^2
		+ 2 \Aa_\cI\, \Aa_\cII\, 
			\t k{_\cI_0} \t k{_\cII_0} \cos(\psi_\cI) \cos(\psi_\cII) \,.
\end{split}
\end{equation}
Now consider the second term in the energy-momentum tensor:
\begin{equation}
	\t \Ff{^\alpha^\beta} \t \Ff{_\alpha_\beta}
		= 4 \Aa_\cI\, \Aa_\cII\, \left[
			g(k_\cI, k_\cII) g(E_\cI, E_\cII)
			- g(k_\cI, E_\cII) g(E_\cI, k_\cII)
		\right] \cos(\psi_\cI) \cos(\psi_\cII)\,.
\end{equation}
By a similar argument as before, one finds this expression to be of order $\varepsilon^2$. Hence, at first order in $\varepsilon$, the energy density reduces to
\begin{equation}
	\t T{_0_0}
		= \left[
			\Aa_\cI\, \t k{_\cI_0} \cos(\psi_\cI)
			+ \Aa_\cII\, \t k{_\cII_0} \cos(\psi_\cII)
		\right]^2\,.
\end{equation}
Finally, we compute the time average of this expression since the read-out cannot resolve the fast oscillations of the electromagnetic field itself but only the slow modulation arising from the gravitational wave.
(Recall that optical frequencies are in the order of $500\,\text{THz}$ while observed \GW\ frequencies are roughly in the range of $10\;\text{to}\;500\,\text{Hz}$). 
Formally, such an average is given by
\begin{align}
	\braket{f(t)}
		& = \frac{1}{T} \int_{t}^{t+T} f(t') \,\dd t'\,,
	&
	&\text{where } \omega \gg T^{-1} \gg \omegaG\,.
\end{align}
In particular, the amplitudes $\Aa$, the polarisation vectors $E$ and the wave vectors $k$ oscillate slowly and may be taken outside the average, leaving us only with the eikonal terms, which oscillate rapidly in time as $- \omega t$. Consequently, we have
\begin{align}
	\braket{\cos(\psi_\cI)^2}
		= \braket{\cos(\psi_\cII)^2}
		&= \half\,,
	&
	\braket{\cos(\psi_\cI) \cos(\psi_\cII)}
		&= \half \cos(\psi_\cI - \psi_\cII)\,,
\end{align}
so that the averaged energy density at the output is
\begin{equation}
	\braket{\t T{_0_0}}
		= \half (\Aa_\cI\, \t k{_\cI_0})^2
		+ \half (\Aa_\cII\, \t k{_\cII_0})^2
		+ (\Aa_\cI\, \t k{_\cI_0}) (\Aa_\cII\, \t k{_\cII_0}) \cos(\psi_\cI - \psi_\cII)
		\,.
\end{equation}

To conclude this section, we verify explicitly that the conditions \eqref{eq:energy inner producs} are satisfied for the fields given in the main body of the text.
Writing \cref{eq:michelson K1,eq:michelson K2} as
\begin{align}
	K_\cI
		&= - \dd t + \nu_\cII
		+ \half \epsilon \alpha [ \omega \Rbs^\o0 \kappa + \mII (\kappa.\kc_\cI)) ]\,,
	\\
	K_\cII
		&= - \dd t + \nu_\cII
		+ \half \epsilon \beta [ \omega \kappa + \mII (\kappa.\kc_\cII)) ]\,,
\end{align}
where the precise form of the factors $\alpha$ and $\beta$ is irrelevant for the argument here, one finds
\begin{equation}
	g(K_\cI, K_\cII)
		= \half \epsilon \alpha \left[
			\omega(- \dd t - \mI).\kappa + \kappa.\kc_\cI
		\right]
		+ \half \epsilon \beta \left[
			\omega(- \dd t - \mII).\kappa + \kappa.\kc_\cII
		\right]
		+ O(\varepsilon^2)
		\,,
\end{equation}
which vanishes since $\kc_\cI = \omega(- \dd t - \mI)$ and $\kc_\cII = \omega(- \dd t - \mII)$.

Similarly, using the expressions for the polarisation vectors from \cref{eq:michelson E1,eq:michelson E2} together with the explicit form of $\delta E^\parallel$ given in \cref{eq:polarisation perturbation decomposition}, as well the fact that $\RI^\o0$, $\RII^\o0$ and $\Rbs^\o0$ are orthogonal matrices, their inner product evaluates to
\begin{equation}
	\begin{split}
		g(E_\cI, E_\cII)
			={}& 1+ 
			\varepsilon \t E{^{\o0}_\mu} [
				\t{\varGamma(\kc_\cI, \kappa.x, \kappa.x^R_\cI)}{^\mu_\nu}
				+ \t{\varGamma(k_\cI, \kappa.x^R_\cI, \kappa.x^E_\cI)}{^\mu_\nu}
			] \t E{^{\o0}^\nu}
			\\&
			+ \varepsilon \t E{^{\o0}_\mu} \t{{\Rbs}}{^{\o0}^\mu_\nu} [
				\t{\varGamma(\kc_\cII,\kappa.x, \kappa.x^R_\cII)}{^\nu_\rho}
				+ \t{\varGamma(k_\cII,\kappa.x^R_\cII, \kappa.x^E_\cII)}{^\nu_\rho}
			] \t{{\Rbs}}{^{\o0}^\rho_\sigma} \varepsilon \t E{^{\o0}^\sigma}
			\\&
			- \half \varepsilon h(E^\o0, E^\o0)_{\kappa.x}
			+ \half \varepsilon h(E^\o0,E^\o0)_{\kappa.x^E_\cI}
			\\&
			- \half \varepsilon h(\Rbs^\o0 E^\o0, \Rbs^\o0 E^\o0)_{\kappa.x}
			+ \half \varepsilon h(\Rbs^\o0 E^\o0, \Rbs^\o0 E^\o0)_{\kappa.x^E_\cII}
			+ O(\varepsilon^2)\,.
	\end{split}
\end{equation}
This reduces to $g(E_\cI, E_\cII) = 1 + O(\varepsilon^2)$ by virtue of the explicit form of the integrated Christoffel symbols given in \cref{eq:Christoffel symbols integrated}.

\clearpage
\markright{References}
\singlespacing\footnotesize
\printbibliography[heading=bibintoc]
\end{document}